\documentclass[twocolumn,superscriptaddress,
amsmath,amssymb,aps,prb]{revtex4-2}

\usepackage{graphicx}
\usepackage{physics}
\usepackage{comment}
\usepackage{color}
\usepackage{dcolumn}
\usepackage{bm}

\usepackage[colorlinks,urlcolor=blue,citecolor=blue,linkcolor=blue]{hyperref}

\newcommand{\up}{\uparrow}
\newcommand{\down}{\downarrow}

\renewcommand{\r}{{\bf r}}
\renewcommand{\k}{{\bf k}}

\newcommand{\q}{{\bf q}}

\newcommand{\eBX}{\varepsilon_{\rm{BX}}}

\newcommand{\nn}{\nonumber}
\newcommand{\beq}{\begin{equation}}
\newcommand{\eeq}{\end{equation}}

\newcommand{\pdag}{{\phantom{\dag}}}
\newcommand{\FS}{\ket{\text{FS}}}

\newcommand{\ham}{\hat{H}}

\newcommand{\area}{\mathcal{A}}

\begin{document}

\title{Multi-dimensional spectroscopy of mobile excitons in two-dimensional semiconductors}

\author{Ned Wheaton}
\affiliation{School of Physics and Astronomy, Monash University, Victoria 3800, Australia}

\author{Jeffrey A. Davis}
\affiliation{Optical Sciences Centre, Swinburne University of Technology, Hawthorn, 3122, Victoria, Australia}

\author{Jesper Levinsen}
\affiliation{School of Physics and Astronomy, Monash University, Victoria 3800, Australia}

\author{Meera M. Parish}
\affiliation{School of Physics and Astronomy, Monash University, Victoria 3800, Australia}

\date{\today}

\begin{abstract}
Multi-dimensional coherent spectroscopy (MDCS) goes beyond standard linear-response probes and provides a powerful tool for investigating correlations between quasiparticles such as excitons (bound electron-hole pairs). Here we present a microscopic theory of MDCS that accounts for the delocalized nature of excitons in two-dimensional semiconductors. In contrast to the more phenomenological few-level approaches typically employed for modelling MDCS, our theory features mobile excitons with continuous momentum degrees of freedom. We find that the energy continuum associated with exciton momenta is crucial for producing interaction-induced decoherence, as well as capturing the interference between different exciton-polaron quasiparticles in the case of charge-doped semiconductors. 
Crucially, our calculated MDCS spectra agree well with recent experiments on doped monolayer MoSe$_2$~[Hao et al.,~Nature Communications \textbf{8}, 15552 (2017)], and they suggest that the interactions between exciton polarons depend strongly on phase-space filling effects, where exciton polarons compete for electrons. 
Our results demonstrate that microscopic approaches allow one to gain new insights from the fine structure of MDCS on two-dimensional semiconductors, and they 
illustrate the utility of microscopic approaches to modelling MDCS experiments more generally. 
\end{abstract}

\maketitle

\section{Introduction}

The development of ultrafast lasers has led to a proliferation of new spectroscopic procedures sensitive to the time-resolved dynamics of solid-state systems, thus offering new insight into the dynamics of correlated many-body systems. Multi-dimensional coherent spectroscopy (MDCS), which uses a series of ultrafast pulses to probe the third-order response of a sample, has proven to be an exceptionally useful tool in the study of many-body physics~\cite{Tollerud2017a,smallwood_multidimensional_2018}. It has seen effective application in two-dimensional (2D) semiconductors such as quantum wells~\cite{Li2006,Stone2009,Turner2010} and transition metal dichalcogenide (TMD) monolayers~\cite{Hao2016,Muir2022,Chen2024,li2023optical}, where the spectra obtained with MDCS bear signatures of excitons (bound electron-hole pairs) and their correlations. Broadly speaking, MDCS is perfectly suited for the study of interactions between several optically accessible excitations.

MDCS experiments are typically modelled using ensembles of isolated few-level systems whose dynamics are computed from some variant of the optical Bloch equations~\cite{li2023optical}. Many-body effects are then usually introduced phenomenologically by adjusting decoherence rates and transition frequencies~\cite{Li2006,smallwood2025}. Although these models can reproduce many experimentally observed features, they offer only limited insight into the underlying physical mechanisms, especially in many-body systems featuring strong correlations. A model which more faithfully describes the microscopic many-body physics at play can do more to reveal the physical 
significance of peak placement, broadening, and intensity. While there are recent efforts to extend the theory of nonlinear response beyond few-level models, these have mainly focused on one-dimensional systems~\cite{Phuc2021,Hart2023,Fava2023,Wang2024,Fava2025}, or the excitations of disordered~\cite{Salvador2024,Salvador2025} or insulating~\cite{Chen2025} systems. There are also recent microscopic theories of exciton quasiparticles in 2D semiconductors~\cite{Tempelaar2019,Hu2024}, but these neglect interactions between quasiparticles, which are known to play a key role in MDCS spectra~\cite{Muir2022}. Thus, a complete microscopic theory of exciton quasiparticles in MDCS is currently lacking.

The behavior of exciton quasiparticles is particularly important for understanding TMD monolayers, which feature strongly bound mobile excitons, a result of reduced screening in 2D materials~\cite{wang2018}. The optical response is thus dominated by excitons and excitonic complexes such as trions (bound states of an exciton and an additional charge carrier) and biexcitons (two-exciton bound states).
With increasing charge doping, these excitons form new quasiparticles known as Fermi polarons~\cite{Sidler2017,Efimkin2017}, which are associated with two branches in the optical spectrum: a lower-energy ``attractive'' polaron and a higher-energy metastable ``repulsive polaron'' adiabatically connected to the trion and exciton, respectively, at low doping. While the properties of individual polarons are now generally well understood~\cite{Massignan2014,Scazza2022,Massignan2026}, the behavior of multiple interacting polarons remains a central concern of recent theoretical efforts~\cite{Guardian2024,Levinsen2025,Levinsen2026}. Observations of polaron-polaron interactions in ultracold gases~\cite{Baroni2024} and in monolayer semiconductors~\cite{Tan2020,Muir2022,Ni2025} offer a wealth of complex data, with differing results depending on the constraints on the mediums and whether the bosonic impurities are quantum degenerate or not~\cite{Levinsen2025,Levinsen2026}. 
In the interest of achieving a full understanding of interacting polarons, high-quality experimental benchmarks are needed; MDCS of exciton polarons in doped TMD monolayers furnishes us with one such benchmark, but its precise import can be unclear in the absence of physically descriptive models.

In this paper, we present a microscopic model of excitons in TMD monolayers that goes beyond previous theories~\cite{Tempelaar2019,Muir2022,Hu2024} by incorporating both quasiparticle interactions and momentum degrees of freedom. We first consider the third-order response for a generic Hamiltonian, using time-dependent perturbation theory to derive expressions for the different MDCS protocols in terms of one- and two-exciton correlation functions. We then consider two scenarios for interacting excitons in undoped 2D semiconductors: one involving immobile excitons and the other featuring excitons with continuous momentum degrees of freedom. 
From the resultant MDCS spectra, we show how features of many-body Hamiltonians manifest in the third-order response, paying particular attention to signatures of interactions. We demonstrate in particular that the inclusion of continuous degrees of freedom allows interaction-induced broadening effects to emerge naturally from the microscopic model. 

Finally, we obtain for the first time a microscopic model of the doping-dependent MDCS spectrum of interacting exciton polarons in doped MoSe$_2$ monolayers. We focus here on co-circular polarization with indistinguishable excitons, and allow the excitons to interact only through an indirect phase-space filling effect driven by competition between polarons for dressing electrons. We obtain remarkably close quantitative agreement with past experiments~\cite{Hao2016,hao2017neutral}. Thus our model serves not only to illustrate the signatures of continuous degrees of freedom in 2D spectra, but also to suggest that phase-space filling is a leading contribution to interactions between bosonic impurities in a Fermi sea.

The paper is organized as follows. In Sec.~\ref{sec:theory}, we derive an expression for the time-dependent nonlinear optical response of a semiconductor. In Sec.~\ref{sec:simple}, we use this expression to simulate MDCS spectra for two model Hamiltonians which describe an undoped 2D semiconductor hosting immobile and mobile excitons, respectively. In Sec.~\ref{sec:doped}, we present a new microscopic model of exciton polarons in doped MoSe$_2$, compare our simulated spectrum to experimental data from Ref.~\cite{hao2017neutral}, and discuss how features of the spectrum relate to features of the underlying microscopic interactions. We conclude and present an outlook in Sec.~\ref{sec:conc}.

\section{Optical response of mobile excitons} \label{sec:theory}

We start by presenting the theoretical formalism for the optical response of a 2D semiconductor such as a TMD monolayer. This follows in large part the standard derivation of the theory of MDCS~\cite{Mukamel2000,hamm2005principles}, but we assume from the outset that we optically excite mobile rather than localized excitons. We also assume, for simplicity, that the system is in a pure state, since we are focused on the microscopic description of semiconductors at low temperatures. However, it is straightforward to go beyond this and consider, for instance, a thermal average over multiple states. 

Since the measured signal in optical spectroscopy is proportional to the induced electric polarization~\cite{Mukamel2000}, we need to calculate the time-dependent expectation value
\begin{align}
P_\sigma(t) = \expval{\hat\mu_\sigma}{\Psi(t)} ,
\end{align}
where $\sigma$ is the spin (circular polarization) of the emitted light, and $\hat{\mu}_\sigma$ is the transition electric dipole moment operator.
Here, the state $\ket{\Psi(t)}$ is time evolving under the combined Hamiltonian $\hat{\mathcal{H}}(t) = \hat{H} + \hat{V}(t)$, where $\hat{H}$ is the time-independent Hamiltonian of the semiconductor and $\hat{V}(t)$ is the small time-dependent perturbation due to the applied optical field:
\begin{align}
\hat{V}(t) = \sum_\sigma \hat{\mu}_\sigma \mathcal{E}_\sigma(t),
\end{align}
with $\mathcal{E}_\sigma(t)$ the (classical) electric field of the applied light pulses for a given polarization $\sigma$. Note that this assumes the dipole approximation for light-matter interactions.

The transition dipole moment operator for mobile excitons with in-plane dipole moment $\mu_X$ takes the form
\begin{align} \label{eq:dipole}
\hat{\mu}_\sigma = \mu_X \left(\hat{X}^\dag_\sigma + \hat{X}_\sigma \right),
\end{align}
where $\hat{X}_\sigma^\dag$ and $\hat{X}_\sigma$ respectively create and annihilate an exciton with spin $\sigma$ and zero in-plane momentum (since the light has negligible in-plane momentum). Here we focus on the lowest energy 1s excitons, but our approach can easily be generalized to include excited Rydberg states. To simplify the notation, we have suppressed the momentum degrees of freedom of the excitons, but we will reinstate these later.
We have also assumed that the transition dipole moment of the exciton is independent of spin, which is reasonable in the absence of a strong magnetic field.

Now, in the interaction picture we have
\begin{align}
 \ket{\Psi(t)} = \hat{U}(t,t_0) \ket{\Psi_I(t)} = e^{-i\hat{H}(t-t_0)} \ket{\Psi_I(t)}, 
\end{align}
where $t_0$ is a time in the distant past before the perturbation is applied. 
Here, and in the following, we set $\hbar=1$.
Then we have 
\begin{align} \label{eq:pol}
P_\sigma(t) & = \bra{\Psi_I(t)}  \underbrace{\hat{U}^\dag(t,t_0) \hat{\mu}_\sigma  \hat{U}(t,t_0)}_{\hat{\mu}_\sigma(t)} \ket{\Psi_I(t)} ,
\end{align}
as well as the transformed light-matter interaction
\begin{align}
\hat{V}_I(t) & = \hat{U}^\dag(t,t_0) \hat{V}(t)  \hat{U}(t,t_0) = \sum_\sigma \hat{\mu}_\sigma(t) \mathcal{E}_\sigma(t).
\end{align}

The state in the interaction picture can be written as an expansion in the perturbation
\begin{align} \nn
 &\ket{\Psi_I(t)}   =  \ket{\Psi(t_0)} - i \int^t_{t_0} d\tau \, \hat{V}_I(\tau)  \ket{\Psi(t_0)} \\ \nn
 &+ i^2 \int^t_{t_0} d\tau_2 \int_{t_0}^{\tau_2} d\tau_1 \, \hat{V}_I(\tau_2) \hat{V}_I(\tau_1)  \ket{\Psi(t_0)} \\ \nn
 & - i^3  \int^t_{t_0} d\tau_3 \int^{\tau_3}_{t_0} d\tau_2 \int_{t_0}^{\tau_2} d\tau_1 \, \hat{V}_I(\tau_3) \hat{V}_I(\tau_2) \hat{V}_I(\tau_1)  \ket{\Psi(t_0)}  \\ \label{eq:psi}
 & + \, \ldots .
\end{align}
Different types of experimental measurements can then isolate and probe different powers of $\hat{V}$ in the expansion for the induced electric polarization.

\begin{figure}
\centering
\includegraphics[width=0.9\columnwidth]{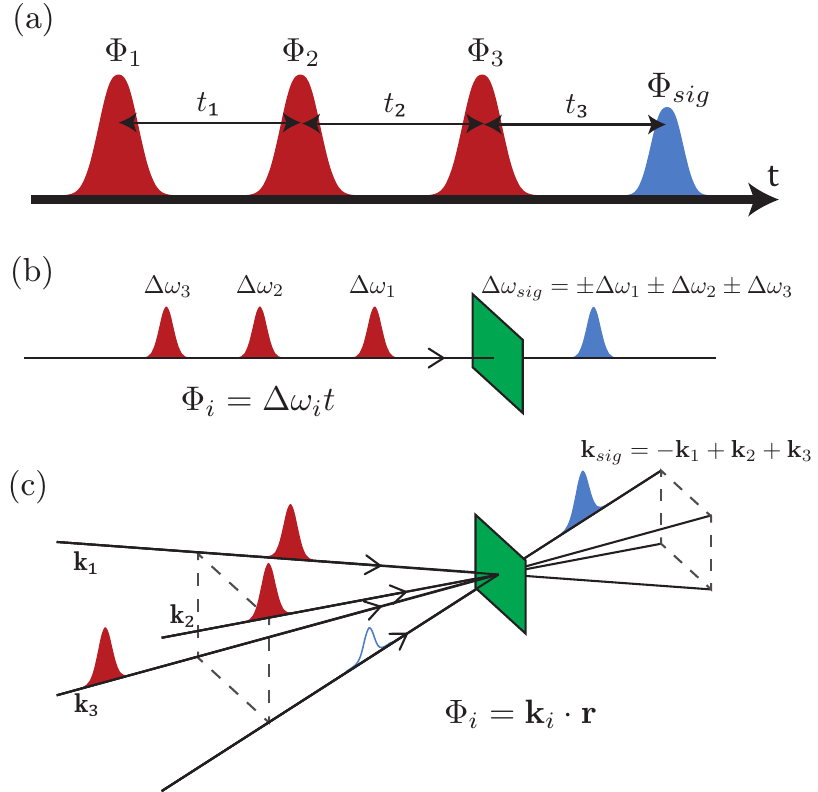}
\caption{Pulse ordering and signal separation. (a) The time ordering of the excitation pulses and delays between them, $t_i$, are defined with each pulse having a phase, $\Phi_i$, which can be defined experimentally in different ways. The signal has a phase given by $\Phi_{sig}=\pm\Phi_1\pm\Phi_2\pm\Phi_3$, where the different sign combinations define the different pathways. (b) In collinear experiments, the phase of each pulse can be given a time dependent phase variation by imparting a small frequency shift, $\Delta\omega_i$. The specific signal pathway can be isolated by selecting the specific beat frequency, $\Delta\omega_{sig}$. (c) In non-collinear (box geometry) configurations, the different phases come from the different wavevectors, $\textbf{k}_i$, and the signal is selected spatially for a specific pathway.}
\label{fig:pulse-seq}
\end{figure}

\subsection{Linear response}
The lowest-order term in the expansion corresponds to linear response, i.e., an induced polarization that depends linearly on the perturbing optical field $\hat{V}$.
Substituting the first term of Eq.~\eqref{eq:psi} into Eq.~\eqref{eq:pol} and taking $t_0 \to -\infty$ thus yields the spin-resolved linear response
\begin{align}
P^{(1)}_{\sigma\sigma'} (t) = -i \int^t_{-\infty} d\tau  \, \mathcal{E}_{\sigma'}(\tau)
\left<\hat{\mu}_\sigma(t) \hat{\mu}_{\sigma'}(\tau) - \hat{\mu}_{\sigma'}(\tau) \hat{\mu}_\sigma(t) \right> ,
\end{align}
where we have defined $\left< \cdots \right> \equiv \expval{\cdots}{\Psi(t_0)}$.
Here, $\sigma'$ and $\sigma$ are the spins (circular polarizations) of the applied and emitted light, respectively. 
Assuming that the initial state $\ket{\Psi(t_0)}$ is the ground state of $\hat{H}$ (i.e., no excitons) and then using the transition dipole moment operator in Eq.~\eqref{eq:dipole}, we obtain
\begin{align} 
P^{(1)}_{\sigma\sigma'} (t) 
 =&  
 \mu_X^2  \int^\infty_{0} dt'  \,\mathcal{E}_{\sigma'}(t-t')
 G_{\sigma\sigma'}(t') 
 \,,
\end{align}
in terms of the single-exciton Green's function
\begin{align}
G_{\sigma\sigma'}(t)  
= -i \Theta(t)\expval*{\hat{X}_\sigma \, e^{-i\hat{H} t} \hat{X}_{\sigma'}^\dag} ,
\end{align}
where $\Theta(t)$ is the Heaviside function. Here, we have set the energy of the ground state to zero, and dropped the complex conjugate term by convention, since this carries no additional information~\cite{Mukamel2000,hamm2005principles}. In the following, we will denote $G_{\sigma}(t) \equiv G_{\sigma\sigma}(t)$ for the diagonal component.

For a light pulse of circular polarization $\sigma$, frequency $\omega$, and 
three-dimensional (3D) wave vector $\k$, the electric field  
is 
\begin{align}
 \mathcal{E}_\sigma(t) \simeq \mathcal{E}_\sigma \delta(t) \left(e^{-i\omega t +i \k\cdot\r} + e^{i\omega t -i \k\cdot\r}  \right) ,
\end{align}
where $\mathcal{E}_\sigma$ is the amplitude and $\r$ is the position in 3D space.
Here, we are taking the semi-impulsive limit where the light pulse is short compared to time scales in the material but long compared to the inverse carrier frequency $1/\omega$, and hence we take it proportional to the Dirac delta function $\delta(t)$. 
If we consider light that is resonant with the exciton energy, $\omega \simeq E_{\rm X}$, and employ the rotating wave approximation, this selects out the term $e^{-i\omega t+i \k\cdot\r}$ term of $\mathcal{E}_\sigma(t)$ such that we finally obtain
\begin{align}
P_{\sigma\sigma'}^{(1)} (t) =  \mu_X^2
\mathcal{E}_{\sigma'} G_{\sigma\sigma'}(t) \,.
\end{align}
Thus, we can directly access the time-dependent single-particle Green's function. Note that the spatial dependence of the electric field varies slowly over the sample (the dipole approximation) and just leads to an overall phase factor in front of the Green's function, which we have suppressed here. 
However, when constructing the response, we need to keep track of the wave vectors (or phase variations) of the applied fields in order to determine which parts of the signal we are selecting.

The Fourier transform of the response is related to the single-exciton spectrum 
since we have 
\begin{align}
A_{\sigma\sigma'}(\omega) = - \frac{1}{\pi} \Im \int^\infty_{-\infty}  dt \, G_{\sigma\sigma'}(t) e^{i\omega t}\,.
\end{align}
When $\sigma = \sigma'$, this corresponds to the spin-resolved spectral function $A_\sigma(\omega) \equiv A_{\sigma\sigma}(\omega)$, which is proportional to absorption. Alternatively, the case where $\sigma \neq \sigma'$ provides information about spin-changing processes, which can be probed  in TMDs by exciting one spin/valley and measuring the emission from the opposite spin/valley.

\subsection{Third-order response}
The lowest non-linear optical response is third order for excitons.  
The second-order response vanishes since it involves correlators with three exciton operators such that we cannot add and remove the same number of excitons. 
In principle, TMD monolayers do feature a second-order optical response for other electronic excitations owing to their lack of inversion symmetry. However, MDCS experiments can filter out any second-order signals since they selectively detect signals that satisfy phase-matching conditions corresponding to the third-order response, as discussed further below.

Using Eq.~\eqref{eq:psi}, we have the third-order spin-resolved response:
\begin{widetext}
\begin{align} \notag
P^{(3)}_{\sigma_4\sigma_3\sigma_2\sigma_1}(t) = & - i^3 \int_{t_0}^{t} d\tau_3 \int_{t_0}^{\tau_3} d\tau_2 \int_{t_0}^{\tau_2} d\tau_1\,
\mathcal{E}_{\sigma_3}(\tau_3) \mathcal{E}_{\sigma_2}(\tau_2) \mathcal{E}_{\sigma_1}(\tau_1) \expval{\hat{\mu}_{\sigma_4}(t)\hat{\mu}_{\sigma_3}(\tau_3) \hat{\mu}_{\sigma_2}(\tau_2) \hat{\mu}_{\sigma_1}(\tau_1)} \\ 
& - i^3 \int_{t_0}^{t} d\tau_3 \int_{t_0}^{t} d\tau_2 \int_{t_0}^{\tau_2} d\tau_1\,
\mathcal{E}_{\sigma_3}(\tau_3) \mathcal{E}_{\sigma_2}(\tau_2) \mathcal{E}_{\sigma_1}(\tau_1) \expval{\hat{\mu}_{\sigma_1}(\tau_1) \hat{\mu}_{\sigma_2}(\tau_2) \hat{\mu}_{\sigma_4}(t)  \hat{\mu}_{\sigma_3}(\tau_3)} + \textrm{h.c.} \,,
\end{align}
where we once again assume that the initial state does not contain any excitons. Taking $t_0 \to -\infty$ and changing variables
to be in terms of the delays between pulses, we have
\begin{equation} \label{eq:P3}
P^{(3)}_{\sigma_4\sigma_3\sigma_2\sigma_1}(t) =  
i \mu_X^4 \int_{0}^{\infty} dt'_3 \int_{0}^{\infty} dt'_2 \int_{0}^{\infty} dt'_1\,
\mathcal{E}_{\sigma_3}(t-t'_3) \mathcal{E}_{\sigma_2}(t-t'_2-t'_3) \mathcal{E}_{\sigma_1}(t-t'_1-t'_2- t'_3) S_{\sigma_4\sigma_3\sigma_2\sigma_1}(t'_3,t'_2,t'_1)\,, 
\end{equation}
where we have defined the response function
\begin{align} \notag
S_{\sigma_4\sigma_3\sigma_2\sigma_1}(t_3,t_2,t_1) 
=   &
\expval{\hat{X}_{\sigma_4} e^{-i\hat{H} t_3} (\hat{X}_{\sigma_3} + \hat{X}_{\sigma_3}^\dag) e^{- i \hat{H} t_2} (\hat{X}_{\sigma_2} + \hat{X}_{\sigma_2}^\dag) e^{-i\hat{H} t_1}\hat{X}_{\sigma_1}^\dag} 
\\ \notag 
& + \expval{\hat{X}_{\sigma_1} e^{i\hat{H} t_1} (\hat{X}_{\sigma_2} + \hat{X}_{\sigma_2}^\dag) e^{i \hat{H} (t_2+t_3)} (\hat{X}_{\sigma_4} + \hat{X}_{\sigma_4}^\dag) e^{-i\hat{H} t_3}\hat{X}_{\sigma_3}^\dag} \\ \notag
& +  \expval{\hat{X}_{\sigma_1} e^{i\hat{H} (t_1+t_2)} (\hat{X}_{\sigma_3} + \hat{X}_{\sigma_3}^\dag) e^{i \hat{H} t_3} (\hat{X}_{\sigma_4} + \hat{X}_{\sigma_4}^\dag) e^{-i\hat{H} (t_2+t_3)}\hat{X}_{\sigma_2}^\dag} \\ 
\label{eq:Sttt}
&  +  \expval{\hat{X}_{\sigma_2} e^{i\hat{H} t_2} (\hat{X}_{\sigma_3} + \hat{X}_{\sigma_3}^\dag) e^{i \hat{H} t_3} (\hat{X}_{\sigma_4} + \hat{X}_{\sigma_4}^\dag) e^{-i\hat{H} (t_1+t_2+t_3)}\hat{X}_{\sigma_1}^\dag}
+ \textrm{h.c.}
\end{align}
Here $\sigma_1$, $\sigma_2$ and $\sigma_3$ are the circular polarizations of the applied optical field, in the order in time in which it interacts with the sample, while $\sigma_4$ is the polarization of the signal. In total, there are 16 non-zero expectation values in $S_{\sigma_4\sigma_3\sigma_2\sigma_1}(t_3,t_2,t_1)$ that cover all the different allowed pathways with equal numbers of creation and annihilation operators.

MDCS experiments separate different components of the third-order response by selecting the specific phase matching conditions of the measured signal. To do this, each pulse is given a unique phase variation, $\Phi_i$, either in space or time that originates from each pulse having a unique wavevector, $\k_i$, or a unique frequency shift in the MHz ($\Delta\omega$), respectively, as depicted in Fig.~\ref{fig:pulse-seq}.
Each component of the signal is then tagged with a specific linear combination of the $\Phi_i$'s.
Throughout this work, we maintain the labelling of each pulse based on the arrival time (i.e., pulse 1 arrives first, pulse 2 second and pulse 3 third). In this case, the different signal components arise for different phase matching conditions, i.e., different linear combinations of the $\Phi_i$'s \footnote{In practice, for experiments done in a non-collinear geometry, where each beam has a different wavevector, the signal direction, and hence phase matching condition, is typically kept constant and the pulse ordering is instead changed to access the different pathways.}.

\subsubsection{Rephasing pathways}
For the \textit{rephasing} pathways, the phase of the extracted signal is $\Phi_{\rm sig} = -\Phi_1 + \Phi_2 + \Phi_3$. For concreteness, we consider the box geometry involving pulses with different wavevectors (see Fig.~\ref{fig:pulse-seq}), in which case the rephasing pathways correspond to selecting the following component of the applied optical electric field:
\begin{align} \label{eq:Efield}
\mathcal{E}(t) &\simeq \mathcal{E}_{\sigma_1}\delta(t+t_1+t_2) e^{i\omega t -i \k_1\cdot\r}
+ \mathcal{E}_{\sigma_2}\delta(t+t_2) e^{-i\omega t +i \k_2\cdot\r} + \mathcal{E}_{\sigma_3}\delta(t) 
e^{-i\omega t +i \k_3\cdot\r} .
\end{align}
Here, within the semi-impulsive limit, we can use the pulse sequence to determine $S_{\sigma_4\sigma_3\sigma_2\sigma_1}(t_3,t_2,t_1)$, where $t_1$, $t_2$, and $t_3$ correspond exactly to the time intervals between pulses, as shown in Fig.~\ref{fig:pulse-seq}. 
Moreover, within the rotating wave approximation, the response in Eq.~\eqref{eq:P3} only picks out a subset of the terms in Eq.~\eqref{eq:Sttt} when we take $\omega \simeq E_{\rm X}$. 
For instance, the first term does not contribute for this particular pulse sequence, while the second and third terms do contribute, as does the Hermitian conjugate of the fourth term. 

Putting this all together, we finally obtain for the rephasing case
\begin{equation}
    P^\mathrm{R}_{\sigma_4\sigma_3\sigma_2\sigma_1}(t_3,t_2,t_1) =  i  \mu_X^4 \mathcal{E}_{\sigma_1}\mathcal{E}_{\sigma_2} \mathcal{E}_{\sigma_3} e^{-iE_{\rm X} t_1} S^{\text{R}}_{\sigma_4\sigma_3\sigma_2\sigma_1}(t_3,t_2,t_1) \, ,
\end{equation}
where we make the time-delays explicit (see Fig.~\ref{fig:pulse-seq}), and the response function is
\begin{align} \notag
    S^{\text{R}}_{\sigma_4\sigma_3\sigma_2\sigma_1}(t_3,t_2,t_1) = & \, \overbrace{\expval{\hat{X}_{\sigma_1} e^{i\hat{H} t_1} \hat{X}^\dag_{\sigma_2} e^{i \hat{H} (t_2+t_3)} \hat{X}_{\sigma_4} e^{-i\hat{H} t_3}\hat{X}^\dag_{\sigma_3}} }^{G^*_{\sigma_2\sigma_1}(t_1) \, G_{\sigma_4\sigma_3}(t_3)}
+  \overbrace{\expval{\hat{X}_{\sigma_1} e^{i\hat{H} (t_1+t_2)} \hat{X}^\dag_{\sigma_3} e^{i \hat{H} t_3} \hat{X}_{\sigma_4} e^{-i\hat{H} (t_2+t_3)}\hat{X}^\dag_{\sigma_2}}}^{G^*_{\sigma_3\sigma_1}(t_1+t_2) G_{\sigma_4\sigma_2}(t_2+t_3)} \\  \label{eq:P3expt}
& - \expval{\hat{X}_{\sigma_1} e^{i\hat{H} (t_1+t_2+t_3)} \hat{X}_{\sigma_4} e^{-i \hat{H} t_3} \hat{X}^\dag_{\sigma_3} e^{-i\hat{H} t_2}\hat{X}^\dag_{\sigma_2}} \, .
\end{align}
The first two terms (dubbed ``ground state bleach'' and ``stimulated emission'', respectively) can be written as products of single-exciton Green's functions, since there is only ever a single exciton present at any time. 
However, the third and final term (``excited state absorption'') includes a time-evolution operator acting on a doubly excited state and it is thus sensitive to interactions between excitons.
Figure~\ref{fig:feynman}(a) shows the standard representation of these different pathways using double-sided Feynman diagrams~\cite{mukamelbook}. We see that there is a direct mapping between these diagrams and the terms in Eq.~\eqref{eq:P3expt} since the placement of the arrows (optical excitations) in Fig.~\ref{fig:feynman}(a) is directly connected to the ordering of operators in the expectation values.

We obtain the frequency-domain response function by holding one time delay constant and then performing a 2D Fourier transform over the remaining time delays~\cite{Tollerud2017a}.
In the \textit{single quantum} (1Q) rephasing protocol, the first delay $t_1$ is scanned while the second delay $t_2$ is held constant to give a response $S^\mathrm{1Q}_{\sigma_4\sigma_3\sigma_2\sigma_1}(\omega_3,t_2,\omega_1)$. The associated energies $\hbar\omega_1$ and $\hbar\omega_3$ can often be interpreted as the absorption energy and emission energy, respectively, though they more correctly describe the energy associated with the coherences in the $t_1$ and $t_3$ time periods. In the \textit{zero quantum} (0Q) rephasing protocol, conversely, the second delay $t_2$ is scanned while the first delay $t_1$ is held constant to give a response $S^\mathrm{0Q}_{\sigma_4\sigma_3\sigma_2\sigma_1}(\omega_3,\omega_2,t_1)$. Here, $\hbar\omega_2$ corresponds to the energy difference between closely spaced states, or the energy of a Raman-excited mode~\cite{yang_isolating_2008,Tollerud2014,Hao2016a,Hao2017,Kolesnichenko2018}. Thus the 1Q response is given by
\begin{align}
        & S^{\text{1Q}}_{\sigma_4\sigma_3\sigma_2\sigma_1}(\omega_3,t_2,\omega_1) = e^{-\Gamma t_2}\int_0^\infty dt_3 \int_0^\infty dt_1 \,e^{i\omega_3 t_3 - \Gamma t_3}  e^{i\omega_1t_1 - \Gamma t_1} 
        S^\mathrm{R}_{\sigma_4\sigma_3\sigma_2\sigma_1}(t_3,t_2,t_1)\, ,\label{eq:Sss'}
\end{align}
and the 0Q response is given by
\begin{align}
        & S^{\text{0Q}}_{\sigma_4\sigma_3\sigma_2\sigma_1}(\omega_3,\omega_2,t_1) = e^{-\Gamma t_1}\int_0^\infty dt_3 \int_0^\infty dt_2 \,e^{i\omega_3 t_3 - \Gamma t_3}  e^{i\omega_2t_2 - \Gamma t_2} 
        S^{\text{R}}_{\sigma_4\sigma_3\sigma_2\sigma_1}(t_3,t_2,t_1)\, .\label{eq:Sss'0Q}
\end{align}
Here we incorporate a radiative recombination rate $\Gamma$ by way of a phenomenological broadening term $e^{-\Gamma (t_1+t_2+t_3)}$~\cite{2DOSreview2014}.

\subsubsection{Double quantum pathways}

We can also consider the \textit{double quantum} (2Q) pulse sequence, which corresponds to $\Phi_{\rm sig} = \Phi_1 + \Phi_2 - \Phi_3$, such that the selected component of the optical field in the box geometry is
\begin{align} \label{eq:Efield2}
E(t) &\simeq \mathcal{E}_{\sigma_1}\delta(t+t_1+t_2) e^{-i\omega t +i \k_1\cdot\r} + \mathcal{E}_{\sigma_2}\delta(t+t_2) e^{-i\omega t +i \k_2\cdot\r} + \mathcal{E}_{\sigma_3}\delta(t) 
e^{i\omega t -i \k_3\cdot\r} .
\end{align}
As above, within the rotating wave approximation the third-order response,
 \begin{equation}
 P^\mathrm{2Q}_{\sigma_4\sigma_3\sigma_2\sigma_1}(t_3,t_2,t_1) =  i  \mu_X^4 \mathcal{E}_{\sigma_1}\mathcal{E}_{\sigma_2} \mathcal{E}_{\sigma_3} e^{iE_{\rm X}(t_1-2t_2)}S^{\text{2Q}}_{\sigma_4\sigma_3\sigma_2\sigma_1}(t_3,t_2,t_1) ,
 \end{equation}
consists of a subset of all processes, leading to the response function
\begin{align}  \label{eq:P3expt2Q}
    S^{\text{2Q}}_{\sigma_4\sigma_3\sigma_2\sigma_1}(t_3,t_2,t_1) & = 
    \expval{\hat{X}_{\sigma_4} e^{-i\hat{H} t_3} \hat{X}_{\sigma_3} e^{-i \hat{H} t_2} \hat{X}^\dag_{\sigma_2} e^{-i\hat{H} t_1}\hat{X}^\dag_{\sigma_1}} 
-  \expval{\hat{X}_{\sigma_3} e^{i\hat{H} t_3} \hat{X}_{\sigma_4} e^{-i \hat{H} (t_2+t_3)} \hat{X}^\dag_{\sigma_2} e^{-i\hat{H} t_1}\hat{X}^\dag_{\sigma_1}} . 
\end{align}
We see that this cannot be decomposed into single-particle Green's functions and this protocol thus provides direct access to the two-exciton properties. Furthermore, we see that the response vanishes in the limit $t_3=0$. The representation in terms of the usual double-sided Feynman diagrams is shown in Fig.~\ref{fig:feynman}(b). To obtain the frequency-domain 2Q response we leave $t_1$ constant (typically at $t_1=0$) and perform a 2D Fourier transform over $t_2$ and $t_3$ to obtain the frequency-domain 2Q response, $S^\mathrm{2Q}_{\sigma_4\sigma_3\sigma_2\sigma_1}(\omega_3,\omega_2,t_1)$. The 2Q energy, $\hbar\omega_2$ corresponds to the energy of the doubly excited state. Including the phenomenological broadening term $e^{-\Gamma(t_1+t_2+t_3)}$, we get
\begin{align} \label{eq:S2Q}
        & S^{\text{2Q}}_{\sigma_4\sigma_3\sigma_2\sigma_1}(\omega_3,\omega_2,t_1) = e^{-\Gamma t_1}\int_0^\infty dt_3 \int_0^\infty dt_2 \,e^{i\omega_3 t_3 - \Gamma t_3}  e^{i\omega_2t_2 - \Gamma t_2} 
        S^{\text{2Q}}_{\sigma_4\sigma_3\sigma_2\sigma_1}(t_3,t_2,t_1) ,
\end{align}
analogous to the 0Q response in Eq.~\eqref{eq:Sss'0Q}.
\end{widetext}

\begin{figure}
\centering
\includegraphics[width=\columnwidth]{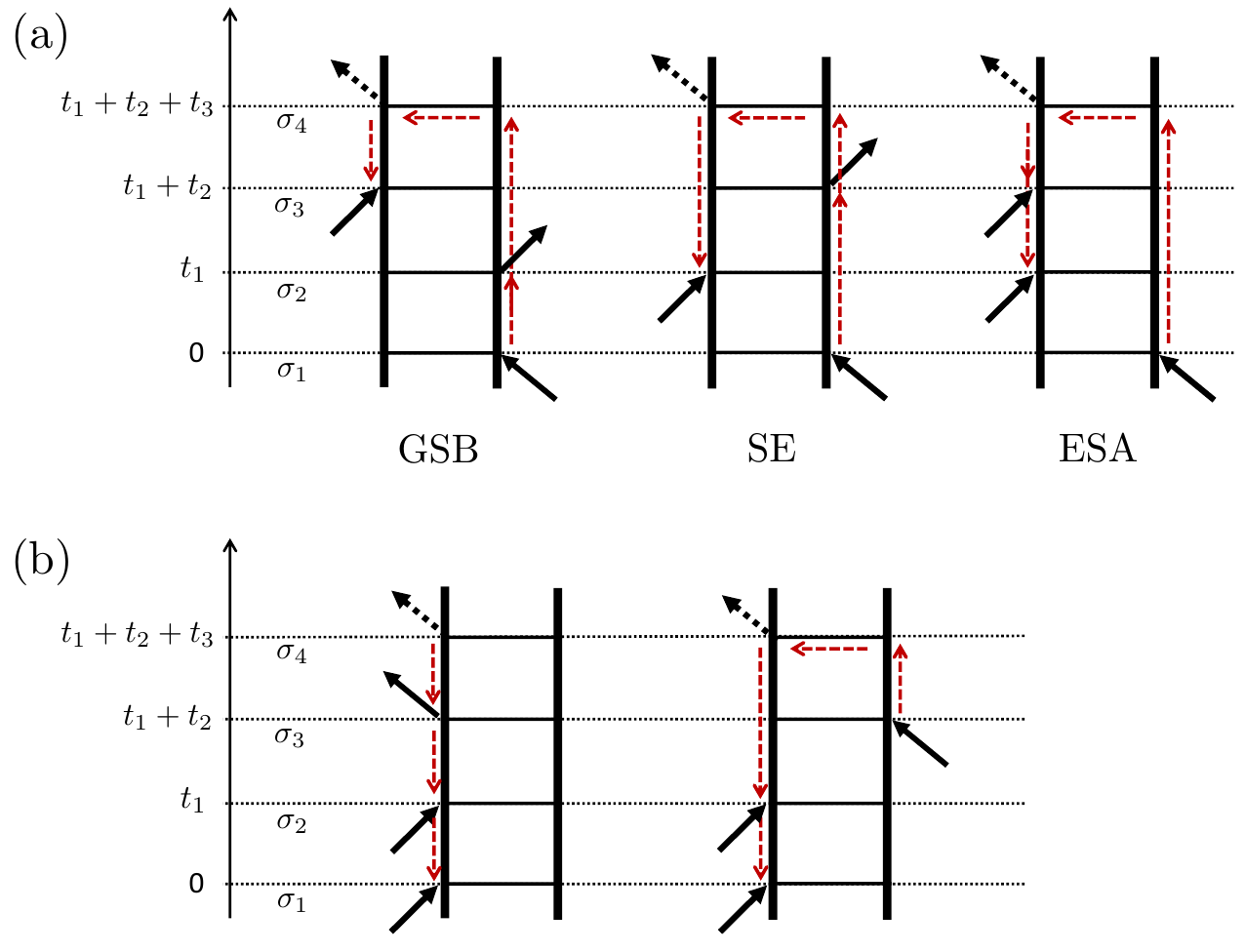}
\caption{Double-sided Feynman diagrams for the different pathways. The rephasing pulse ordering (a) involves three terms  in Eq.~\eqref{eq:P3expt} corresponding to ground-state bleach (GSB), stimulated emission (SE) and excited state absorption (ESA), while the double quantum pulse sequence (b) involves the two terms in Eq.~\eqref{eq:P3expt2Q}. Black arrows pointing to the right (left) represent electric fields with $e^{-i\omega t}$ ($e^{i\omega t}$) and correspond to exciton creation (annihilation) operators.  
The left-to-right ordering of operators in each expectation value in Eqs.~\eqref{eq:P3expt} and \eqref{eq:P3expt2Q} is obtained by traversing each diagram in a counterclockwise loop, as illustrated by the red arrows. The overall sign of each diagram is $(-1)^n$, with $n$ the number of interactions on the right-hand side of the diagram~\cite{mukamelbook}.}
\label{fig:feynman}
\end{figure}

\subsubsection{Co-circular and cross-circular protocols}

In the remainder of this paper, we will limit our attention to sequences $\sigma_4\sigma_3\sigma_2\sigma_1=\sigma\sigma'\sigma\sigma'$ which correspond to the experimentally relevant co-circular ($\up\up\up\up$ and $\down\down\down\down$) and cross-circular ($\uparrow\downarrow\uparrow\downarrow$ and $\downarrow\uparrow\downarrow\uparrow$) polarization protocols. Thus, we adopt the more compact notation $\sigma\sigma' \equiv \sigma\sigma'\sigma\sigma'$ for the response functions.

\subsection{Disorder averaging}

In practice, the exciton energy is not uniform throughout the material; owing to impurities, defects, strain, and other sources of disorder, any real sample contains a distribution of exciton energies. We therefore include inhomogeneous broadening in our model by averaging over such a distribution, which we take to be Gaussian. Its variance, $\gamma$, controls the inhomogeneous linewidth and sets the strength of the disorder. Thus, in the case of a disordered material, a response function such as Eq.~\eqref{eq:Sss'} becomes
\beq
    \langle S^\mathrm{1Q}_{\sigma\sigma'}(\omega_3,t_2,\omega_1) \rangle = \int d\varepsilon\, \frac{e^{-(\varepsilon-E_{\rm X})^2/2\gamma^2}}{\sqrt{2\pi\gamma^2}} S^\mathrm{1Q}_{\sigma\sigma'}(\omega_3,t_2,\omega_1) \,. \label{eq:inhom}
\eeq

\section{Mobile versus localized excitons}
\label{sec:simple}
We first consider models of interacting excitons at low energies, allowing us to derive analytic expressions and thus gain insight into the third-order optical response. We focus in particular on the impact of the biexciton---corresponding to a bound state of an $\up$ and $\down$ exciton---and investigate how the optical response depends on whether the excitons are localized or mobile. 

\subsection{Localized excitons} \label{subsec:immobile} 
We begin with the case of excitons that have been localized due to, e.g., an array of trapping potentials such as in a Moir\'e superlattice~\cite{Brem2020,Dandu2022}. The corresponding Hamiltonian involves immobile excitons with on-site interactions:
\begin{multline}\label{eq:Hsimple}
\ham = \sum_{\sigma} E_{\rm{X}} \hat{X}_{\sigma}^\dag \hat{X}_{\sigma}^\pdag + \frac{U}{2}  \sum_{\sigma} \hat{X}_{\sigma}^\dag \hat{X}_{\sigma}^\dag \hat{X}_{\sigma}^\pdag \hat{X}_{\sigma}^\pdag \\
+ (2E_{\rm{X}} 
- \varepsilon_{\rm{BX}}) \hat{B}^\dag \hat{B} + \lambda (\hat{B}^\dag \hat{X}_{\down}^\pdag \hat{X}_{\up}^\pdag + \hat{X}_{\up}^\dag \hat{X}_{\down}^\dag \hat{B})  ,
\end{multline}
where $E_{\rm{X}}$ is again the exciton energy.  
Here, $U$ is the strength of same-spin exciton-exciton interactions, which are repulsive in TMD monolayers due to the Pauli exclusion between identical electrons and holes. 
By contrast, excitons with opposite spin can form biexcitons (bound states between two excitons)~\cite{Sie2015,hao2017neutral}.
Thus, we explicitly include a coupling to the biexciton state in the Hamiltonian, with creation operator $\hat{B}^\dag$, biexciton binding energy $\varepsilon_{\rm{BX}}$, and coupling strength $\lambda$ (which is chosen to be real and positive).  
Note that, for simplicity, we have focused here on systems with spin/valley symmetry where the exciton properties are independent of spin, but it is straightforward to generalize the results beyond this case. 

We can immediately compute the rephasing response for both co-circular ($\up\up$) and cross-circular ($\up\down$)  polarizations by inserting $\ham$ into Eq.~\eqref{eq:P3expt}, yielding
\begin{subequations}
\begin{align}
    S^\mathrm{R}_{\up\up
    }(t_3,t_2,t_1) =& \, 2 e^{iE_{\rm{X}} (t_1 - t_3)} (1 - e^{-i U t_3}) ,\\  \notag
    S^\mathrm{R}_{\up\down
    }(t_3,t_2,t_1) = & \,  e^{iE_{\rm{X}} (t_1 - t_3)} \\
   & \times (1 - u^2 e^{-i E_+ t_3} - v^2 e^{-i E_- t_3}) \,,
\end{align}
\end{subequations}
where we use the fact that the single-exciton Green's function is diagonal, i.e.,  $G_{\sigma\sigma'}(t)=-i\Theta(t)e^{-iE_\mathrm{X}t}\delta_{\sigma\sigma'}$, with  the Kronecker delta $\delta_{\sigma\sigma'}$. The energies $E_\pm$ (measured from $E_\mathrm{X}$) and coefficients $u$, $v$ 
appearing in the cross-circular response are given by
\begin{subequations} \label{eq:BXeqs}
\begin{align}
    E_\pm &= \frac{1}{2}\left(-\varepsilon_{\rm{BX}} \pm \sqrt{\varepsilon_{\rm{BX}}^2 + 4\lambda^2}\right) ,\\
    u^2 &= 1-v^2 = \frac{E^2_-}{E_-^2 + \lambda^2} \,.
\end{align}
\end{subequations}
In the limit where the coupling $\lambda \ll \eBX$, the lower state with $E_- \simeq -\eBX -\frac{\lambda^2}{\eBX}$ evolves into the biexciton bound state, while the upper state, $E_+ \simeq \frac{\lambda^2}{\eBX}$, corresponds to an unbound state  where the two excitons repel each other.

We see that the rephasing response function $S^\mathrm{R}_{\sigma\sigma'}$ vanishes whenever the excitons are non-interacting, i.e., for either $U=0$ or $\lambda = 0$, the single-particle contributions [ground-state bleach and stimulated emission, the first two terms in Eq.~\eqref{eq:P3expt}] exactly cancel with the two-particle contribution (excited-state absorption, the third and final term). Thus, any non-zero signal in the rephasing spectrum will be a signature of exciton-exciton interactions. 
Furthermore, the response is independent of $t_2$ in this model because the single-exciton states are degenerate in energy. 

\begin{figure}
\centering
\includegraphics[width=\columnwidth]{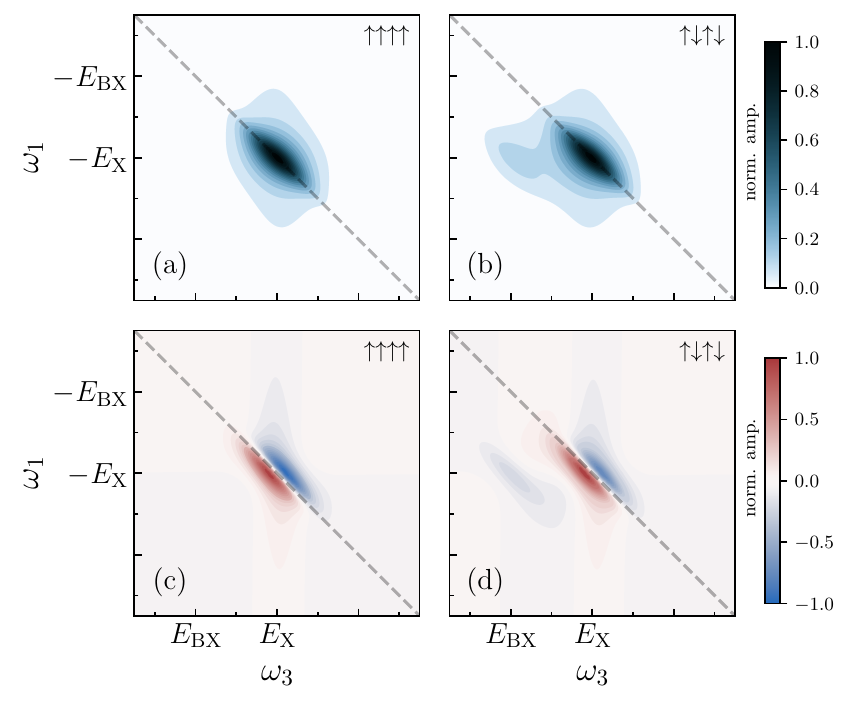}
\caption{Absolute value (a-b) and real part (c-d) of 1Q rephasing spectra at $t_2=0$ for immobile excitons described by the Hamiltonian in Eq.~\eqref{eq:Hsimple} with homogeneous broadening $\Gamma/\varepsilon_{\rm{BX}}=0.1$, and inhomogeneous broadening $\gamma/\varepsilon_{\rm{BX}}=0.2$.  We have used the biexciton binding energy $\varepsilon_{\rm{BX}}$ as a common scale to compare the spectra and we have introduced the notation $E_{\rm{BX}} \equiv E_{\rm X} - \varepsilon_{\rm{BX}}$, the energy needed to create a biexciton starting from a system with a single exciton. The dashed line indicates the diagonal $\omega_3=-\omega_1$. Panels (a) and (c) show co-circular polarization with interaction $U/\varepsilon_{\rm{BX}}=0.01$, while  
panels (b) and (d) show cross-circular polarization with coupling $\lambda/\varepsilon_{\rm{BX}}=0.01$. The maximum values in all spectra are normalized to 1.}
\label{fig:simple}
\end{figure}

To analyze the spectrum, we consider the 1Q rephasing response functions in the frequency domain with phenomenological broadening:
\begin{widetext}
\begin{subequations}
\label{eq:simplespecboth}
\begin{align} \label{eq:simplespec-co}
S^{\text{1Q}}_{\up\up}(\omega_3,t_2,\omega_1) &=\frac{2e^{-\Gamma t_2}}{\omega_1 + E_{\rm{X}} + i\Gamma}  \left( \frac{1}{\omega_3 - E_{\rm{X}} + i\Gamma} - \frac{1}{\omega_3 - (E_{\rm{X}} + U) + i\Gamma} \right)\,,\\ \label{eq:simplespec-cross}
S^{\text{1Q}}_{\uparrow\downarrow}(\omega_3,t_2,\omega_1) &= \frac{e^{-\Gamma t_2}}{\omega_1 + E_{\rm{X}} + i\Gamma} \left( \frac{1}{\omega_3 - E_{\rm{X}} + i\Gamma} - \frac{u^2}{\omega_3 - (E_{\rm{X}} + E_{+}) + i\Gamma} - \frac{v^2}{\omega_3 - (E_{\rm{X}} + E_- ) + i\Gamma}\right) \, .
\end{align}
\end{subequations}
\end{widetext}
Examples of the resulting 2D spectra are shown in Fig.~\ref{fig:simple}. In the co-circular case, a weak repulsive interaction $0<U\ll\Gamma$ leads to a single peak in the absolute value $|S^{\text{1Q}}_{\up\up}|$, as shown in Fig.~\ref{fig:simple}(a), resulting from the imperfect cancellation of two Lorentzian peaks with identical broadening and absorption frequencies $\omega_1=E_{\rm X}$ but slightly different emission frequencies $\omega_3$. Along $\omega_3$, this peak is centred at $\omega_3 \simeq E_{\rm X} + U/2$ and has a full width at half maximum (FWHM)  of $2\Gamma+U^2/2\Gamma + \mathcal{O}(U^4/\Gamma^3)$. Furthermore, its amplitude is $2U/\Gamma^3 + \mathcal{O}(U^2/\Gamma^4)$, which correctly implies that the signal vanishes for $U=0$, as we have emphasized above.
This peak possesses a ``dispersive'' lineshape, meaning that the real part does not resemble a Lorentzian peak centred at emission frequency $\omega_3=E_{\rm X} + U/2$ (which would be an ``absorptive lineshape'') but instead exhibits a sign change at this emission frequency, leading to negative and positive lobes on either side of the resonance. This feature is clearly visible in Fig.~\ref{fig:simple}(c). Note that the direction of this sign change in $\Re[S^{\text{1Q}}_{\up\up}]$ depends on the sign of the interaction; if we instead had an attractive interaction $U<0$, the real part would be negative at lower $\omega_3$ and positive at higher $\omega_3$.  

We can describe this sign change in terms of the complex phase $\theta_{\up\up}$, where the total complex response is given by $S^{\text{1Q}}_{\up\up}=\lvert S^{\text{1Q}}_{\up\up} \rvert e^{i\theta_{\up\up}}$. Dispersive peaks like that seen in Fig.~\ref{fig:simple}(c) exhibit a $\pi$ phase shift across their centres as the emission frequency $\omega_3$ is varied from one side of the central resonance to the other. The same dispersive lineshapes have been seen to characterize the peaks corresponding to heavy-hole excitons in GaAs quantum wells~\cite{Li2006,zhang2007}, though their phase shift is typically less than $\pi$ owing to dephasing processes induced by exciton-exciton interactions~\cite{smallwood2025}. This ``excitation-induced dephasing'' is frequently modelled phenomenologically by adding an imaginary part to the interaction $U$, but we will see below in our discussion of mobile excitons how it can arise naturally from the inclusion of continuous degrees of freedom in our model.
 
For stronger repulsive interactions $U\ge\Gamma$, the individual single-exciton and two-exciton contributions can be resolved in $|S^{\text{1Q}}_{\up\up}|$, giving two peaks of equal amplitude centred at $E_{\rm X}$ and $E_{\rm X}+U$, respectively. Taking the limit $U\to\infty$, we obtain a hard-core repulsion allowing only one exciton per site, leaving a single Lorentzian peak exactly on the diagonal at $\omega_3= -\omega_1 = E_{\rm X}$ with no dispersive lineshape. This is equivalent to the ``Pauli-blocking'' and ``phase-space filling'' models known to inadequately describe semiconductor quantum wells~\cite{Li2006,zhang2007,smallwood_multidimensional_2018,smallwood2025}.

The cross-circular spectrum typically features two clearly distinct peaks in $|S^{\text{1Q}}_{\up\down}|$, as shown in Fig.~\ref{fig:simple}(b) for $\lambda\ll\Gamma\ll\varepsilon_{\rm{BX}}$. We may identify these with the biexciton (emission from $\omega_3 = E_{\rm X} + E_- = E_{\rm X} - \varepsilon_{\rm{BX}} + \mathcal{O}(\lambda^2)$) and the unbound exciton (emission from $\omega_3 = E_{\rm X} + E_+ = E_{\rm X} + \mathcal{O}(\lambda^2)$). The unbound peak is a dispersive peak analogous to that found in the co-circular spectrum in panels (a) and (c), whereas the biexciton peak is a simple Lorentzian peak which originates purely from the two-exciton contribution. This can be seen most clearly in the real part shown in Fig.~\ref{fig:simple}(d), where the unbound peak exhibits a sign change but the biexciton peak is purely negative. Precisely the same behavior has been observed in cross-circular spectra in GaAs quantum wells~\cite{Bristow2009}. Referring to Eq.~\eqref{eq:BXeqs}, we see that the strength of the coupling $\lambda$ controls the amplitude and position of the exciton and biexciton resonances. An increase in the coupling leads to a blueshift in the emission frequency of the exciton peak and an equal redshift in the biexciton peak. In particular, for small coupling $\lambda$ these shifts are given by $(\Delta\omega_3)_\pm\simeq\pm\lambda^2/\varepsilon_{\rm{BX}}$. This is accompanied by a shift in weight from the exciton peak to the biexciton peak, with $v^2 = 1-u^2 \simeq \lambda^2/\varepsilon_{\rm{BX}}^2$ in Eq.~\eqref{eq:simplespec-cross}. The unbound exciton peak, resulting as it does from the imperfect cancellation of the one-exciton and two-exciton contributions, is actually strengthened by this reduction in $u^2$; thus the intensity of both peaks is proportional to $\lambda^2/\varepsilon_{\rm{BX}}$. In principle, the biexciton-exciton coupling $\lambda$ in this model can be enhanced by increasing the depth of the trapping potential for the excitons, i.e.,  reducing the trapping length scale, since this increases the overlap between biexciton and exciton wave functions.

The properties of exciton-exciton correlations are also contained in the double quantum response in Eq.~\eqref{eq:P3expt2Q}, which is straightforward to compute within this model; for completeness, the time-domain and frequency-domain response functions are shown in Appendix~\ref{sec:2Qappendix}.
Once again, a nonzero response indicates the presence of exciton-exciton interactions, and the cross-circular response probes the properties of the biexciton. However, in this case, the co-circular response vanishes for large repulsion $U \to \infty$ since all peaks disappear off to infinity.

\subsection{Mobile excitons} \label{subsec:mobile}

We now consider a more realistic treatment of the microscopic physics governing exciton dynamics, where we include the continuous momentum degrees of freedom associated with mobile excitons. In this case, 
the exciton operator $\hat{X}_{\k\sigma}$ and associated Green's function 
acquires a momentum degree of freedom, such that $G_{\sigma\sigma'}(t,\k) = -i \Theta(t) \expval{\hat{X}^{\pdag}_{\k\sigma} e^{-i\hat{H}t} \hat{X}^\dag_{\k\sigma'}}$. 
The system of interacting excitons can be described by the Hamiltonian:
\begin{multline} \label{eq:Hmobile}
    \ham = \sum_{\k\sigma} (E_{\rm X} + \epsilon^{\rm X}_{\k}) \hat{X}^\dag_{\k\sigma} \hat{X}_{\k\sigma}^\pdag \\ 
    +  \frac{1}{2\mathcal{A}}\sum_{\substack{\k\k'\q \\ \sigma\sigma'}}V_{\sigma\sigma'}(\q) \hat{X}^\dag_{\k+\q\sigma} \hat{X}^\dag_{\k'-\q\sigma'} \hat{X}_{\k'\sigma'}^\pdag \hat{X}_{\k\sigma}^\pdag  . 
\end{multline}
Here, the first term $\hat{H}_0$ contains the kinetic part, where the exciton dispersion $\epsilon^{\rm X}_\k =|\k|^2/2m_{\rm X} \equiv  k^2/2m_{\rm X}$ with exciton mass $m_{\rm X}$, while the second term $\hat{V}$ describes the exciton-exciton interactions, where $V_{\sigma\sigma'}(\q)$ is the short-ranged exciton-exciton interaction potential, and $\mathcal{A}$ is the area of the 2D space containing our excitons.

To calculate the rephasing response in this model, we first note that the single-exciton Green's function is simply $G_{\sigma\sigma'}(t,\k)\equiv G_{\sigma}(t,\k) \delta_{\sigma\sigma'} = -i \Theta(t) e^{-i(E_{\rm X} + \epsilon^{\rm X}_{\k}) t}\delta_{\sigma\sigma'}$, like for the Hamiltonian considered above in Eq.~\eqref{eq:Hsimple}.
Furthermore, the two-exciton ``excited-state absorption'' term in Eq.~\eqref{eq:P3expt} corresponds to a modified exciton Green's function at zero momentum 
\begin{multline} \label{eq:modGF}
    \expval{\hat{X}^\pdag_{0\sigma'} e^{i\hat{H} (t_1+t_2+t_3)} \hat{X}^\pdag_{0\sigma} e^{-i \hat{H} t_3} \hat{X}^\dag_{0\sigma'} e^{-i\hat{H} t_2}\hat{X}^\dag_{0\sigma}} =  \\
   e^{iE_{\rm X}t_1} \underbrace{\expval{\hat{X}^\pdag_{0\sigma'} e^{i\hat{H}t_3} \hat{X}^\pdag_{0\sigma} e^{-i \hat{H} t_3} \hat{X}^\dag_{0\sigma} \hat{X}^\dag_{0\sigma'}}}_{i G^{(\sigma')}_{\sigma}(t_3)} \, ,
\end{multline}
 which describes the dynamics of a spin-$\sigma$ exciton in the presence of another zero-momentum exciton of spin $\sigma'$.
Transforming to frequency space, the Green's function can then be written in terms of a self-energy $\Sigma_{\sigma\sigma'}(\omega)$~\cite{fetterbook}, such that we have
$G_{\sigma}^{(\sigma')}(\omega) = [\omega - E_{\rm X} - \Sigma_{\sigma\sigma'}(\omega)]^{-1}$.
Thus, putting this all together, we finally obtain:
\begin{widetext}
\beq
S^{\text{1Q}}_{\sigma\sigma'}(\omega_3,t_2,\omega_1) = \frac{(1+\delta_{\sigma\sigma'} )e^{-\Gamma t_2}}{\omega_1 + E_{\rm X} + i\Gamma} \left( \frac{1}{\omega_3 - E_{\rm X} +  i\Gamma} - \frac{1}{\omega_3 - E_{\rm X} - \Sigma_{\sigma\sigma'}(\omega_3  +i\Gamma) + i\Gamma} \right) .
\label{eq:selfEresponse}
\eeq
\end{widetext}

The effects of exciton-exciton interactions are contained entirely within the self-energy $\Sigma_{\sigma\sigma'}(\omega)$. Its real part leads to an energetic shift in the exciton resonance, while its imaginary part leads to increased dephasing. Comparing Eq.~\eqref{eq:selfEresponse} with Eq.~\eqref{eq:simplespecboth} above, we see that the system of immobile excitons has a purely real two-exciton self-energy:
\beq 
\Sigma_{\sigma\sigma'}(\omega) = \begin{cases} 
U, & \sigma=\sigma'\\
\lambda^2/(\omega - E_{\rm X} + \varepsilon_{\rm{BX}}) , & \sigma\ne \sigma'
\end{cases}
\eeq
This means that the model of immobile excitons gives rise to energetic shifts and to the appearance of a new biexciton peak in the cross-circular polarization, but it cannot induce dephasing unless one includes an imaginary component in $U$ which introduces another fitting parameter. Conversely, as we demonstrate in the following, the inclusion of continuous momentum degrees of freedom naturally leads to a complex self-energy where the full range of many-body effects, both shifting and dephasing, are present. Crucially, this happens without the need to introduce additional parameters and allows for the modelling of peaks beyond the simple dispersive lineshapes observed in Fig.~\ref{fig:simple}. 

For mobile excitons governed by the Hamiltonian in Eq.~\eqref{eq:Hmobile}, it is straightforward to obtain the self-energy $\Sigma_{\sigma\sigma'}(\omega)$, since it simply corresponds to a two-exciton scattering problem, as can be seen from our definition of the modified Green's function in Eq.~\eqref{eq:modGF}. It follows that all of the relevant physics is contained in the two-body scattering $T$ matrix, $\hat{T}$, and the self-energy takes the form~\footnote{A similar approach has been used to analyze exciton-exciton interactions in quantum wells~\cite{Schindler2008}.} 
\begin{equation}\label{eq:SEss'}
    \Sigma_{\sigma\sigma'}(\omega) = 
    \frac{\expval*{ \hat{X}_{0\sigma'}^\pdag  \hat{X}_{0\sigma}^\pdag \hat{T}(\omega)\hat{X}^\dag_{0\sigma} \hat{X}^\dag_{0\sigma'}}}{1+\delta_{\sigma\sigma'}} \equiv \frac{1}{\mathcal{A}} T_{\sigma\sigma'}(\omega)\,. 
\end{equation}
Here, the area $\mathcal{A}$ in which the $\sigma$ and $\sigma'$ excitons reside is set by the exciton density, which is in turn proportional to the intensity of the light pulses.

We solve for the exciton $T$ matrix by way of the Lippmann-Schwinger equation~\cite{fetterbook},
\beq
\hat{T}(\omega) = \hat{V} + \hat{V}\frac1{\omega-\hat{H}_0+i0} \hat{T}(\omega)\,.
\eeq
Here the factor $+i0$ shifts the poles in $\omega$ slightly into the lower half of the complex plane, which physically ensures that we are considering an outgoing scattered wave~\cite{fetterbook}.
Defining the more general momentum-dependent $T$ matrix $T_{\sigma\sigma'}(\q,\omega) = \frac{\mathcal{A}}{1+\delta_{\sigma\sigma'}} \expval*{\hat{X}_{-\q\sigma'}^\pdag  \hat{X}_{\q\sigma}^\pdag \hat{T}(\omega)\hat{X}^\dag_{0\sigma} \hat{X}^\dag_{0\sigma'}}$, we obtain the integral equation
\beq \label{eq:Tmat}
T_{\sigma\sigma'}(\q,\omega) = V_{\sigma\sigma'}(\q) + \frac{1}{\mathcal{A}}\sum_{\k} \frac{V_{\sigma\sigma'}(\q-\k)}{\omega-E_{\rm X}-2\epsilon_\k} T_{\sigma\sigma'}(\k,\omega) \, .
\eeq
Since excitons in TMDs are created with negligible in-plane momentum such that their kinetic energy will be small compared to the exciton binding energy $\varepsilon_{\rm X}$, i.e., such that $\lvert \omega - E_{\rm X} \rvert \ll \varepsilon_{\rm X}$, we expect the optical response to be governed mainly by low-energy scattering. We therefore take the low-energy limit, in which the exciton wavelength greatly exceeds the interaction range and thus scattering becomes approximately isotropic. The $T$ matrix then assumes the following logarithmic form~\cite{Landau2013}:
\beq \label{eq:mobileselfenergy}
T_{\sigma\sigma'}(\omega) \equiv 
T_{\sigma\sigma'}(0,\omega)= \frac{4\pi}{m_{\rm X}}
\frac1{\ln\left(\frac{\varepsilon_{\sigma\sigma'}}{\omega-E_{\rm X}
}\right)+i\pi}\,.
\eeq
The strength of the interaction is controlled by the parameter $\varepsilon_{\sigma\sigma'}$ which can be measured experimentally. While it is not known exactly for excitonic systems, it can be estimated to logarithmic accuracy~\cite{Bleu2020b}: 
For cross-circular polarization with $\sigma \ne \sigma'$, the energy that sets the scale of the interactions is the biexciton binding energy, i.e., we can take $\varepsilon_{\sigma\sigma'}\simeq \varepsilon_{\rm{BX}}$. In the co-circular case $\sigma=\sigma'$, the exciton-exciton interactions are purely repulsive due to Pauli exclusion and the relevant energy scale is instead the exciton binding energy itself; thus we use the approximation $\varepsilon_{\up\up}=\varepsilon_{\down\down}\simeq\varepsilon_{\rm X}$~\footnote{We note that in the co-circular case, this approach leads to a pole of the two-body $T$ matrix at the exciton energy. This pole is outside the regime of validity of the low-energy expansion, and should be ignored.}.

Importantly, the $T$ matrix, and thus the self-energy, has a non-zero imaginary part whenever $\omega>E_{\rm X}$ which arises from the two-exciton scattering continuum, whereas it cancels with the imaginary part from the logarithm whenever $\omega<E_\mathrm{X}$. While the interaction between immobile excitons, considered above, merely shifted the resonance, we now see that interactions between mobile excitons can broaden resonances as well, giving rise to the ``excitation-induced dephasing'' commonly observed in experiments~\cite{Moody2015}. We can then obtain the 1Q rephasing spectrum from Eq.~\eqref{eq:selfEresponse}.

\begin{figure}
\centering
\includegraphics[width=\columnwidth]{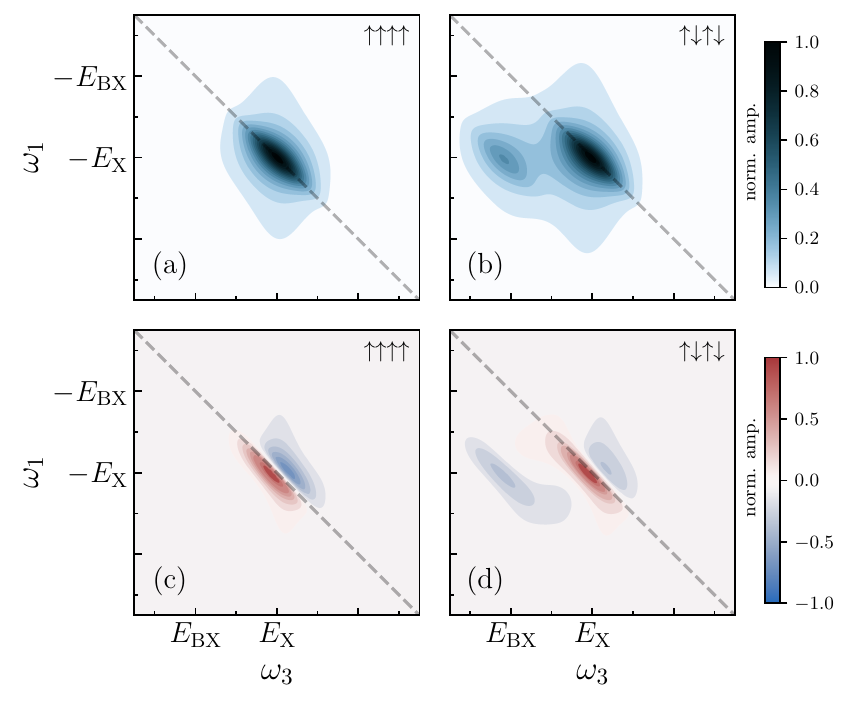}
\caption{Absolute value (a-b) and real part (c-d) of 1Q rephasing spectra at $t_2=0$ for mobile excitons described by the Hamiltonian in Eq.~\eqref{eq:Hmobile}. We show results for the co-circular polarization (a,c), and the cross-circular polarization (b,d), where we take $\varepsilon_{\rm X}=400$ meV and $\varepsilon_{\rm{BX}}=20$ meV.  All spectra are computed with area $\mathcal{A}\cdot m_{\rm X}\varepsilon_{\rm{BX}}=100$, corresponding to an exciton density of $5.25\times10^{11}$ cm$^{-2}$. We also use a homogeneous broadening $\Gamma/\varepsilon_{\rm{BX}}=0.1$, and an inhomogeneous broadening $\gamma/\varepsilon_{\rm{BX}}=0.2$. The maximum values in all spectra are normalized to 1.}
\label{fig:mobile}
\end{figure}

In anticipation of the approach that we will employ in the case of doped semiconductors in Sec.~\ref{sec:doped}, we now show that we can obtain the same two-exciton $T$ matrix by considering the general two-exciton state,
\begin{multline}
\hspace{-3mm}    \ket{\psi_{\sigma\sigma'}} = \\ \frac1{\sqrt{1+\delta_{\sigma\sigma'}}}\left(\alpha_0 \hat{X}^\dag_{0\sigma} \hat{X}^\dag_{0\sigma'} + \frac{1}{\sqrt{\mathcal{A}}} \sum_\k \alpha_\k \hat{X}^\dag_{\k\sigma} \hat{X}^\dag_{-\k\sigma'}\right) \ket{0}
    ,
\end{multline}
which includes all two-body scattering states with zero total momentum. Here, the amplitudes $\alpha\in\{\alpha_0,\alpha_\k\}$ are treated as variational parameters, and we suppress their spin dependence. We obtain equations of motion by requiring $\partial_{\alpha^*} \bra{\psi_{\sigma\sigma'}} (\ham - E) \ket{\psi_{\sigma\sigma'}} = 0$, which yields
\begin{subequations} \label{eq:vartwo}
\begin{align}
    (E-2E_{\rm X})\alpha_0 &= \frac{V_{\sigma\sigma'}(0)}{\mathcal{A}} \alpha_0 + \sum_{\k'} \frac{V_{\sigma\sigma'}(\k')}{\mathcal{A}^{3/2}} \alpha_{\k'}\,, \label{eq:vartwo1}\\
    (E-2E_{\rm X}-2\epsilon_\k)\alpha_\k &= \frac{V_{\sigma\sigma'}(\k)}{\mathcal{A}^{1/2}} \alpha_0 + \sum_{\k'} \frac{V_{\sigma\sigma'}(\k-\k')}{\mathcal{A}} \alpha_{\k'}
    \label{eq:vartwo2}
    \,.
\end{align}
\end{subequations}
The energy $E$ of the two-exciton system is related to the frequency $\omega$ in Eq.~\eqref{eq:SEss'} via $E=\omega+E_\mathrm{X}$, since the self-energy $\Sigma_{\sigma\sigma'}(\omega)$ is evaluated for a spin-$\sigma$ exciton in the presence of a $\sigma'$ exciton. Furthermore, by inserting Eq.~\eqref{eq:vartwo2} into Eq.~\eqref{eq:vartwo1} and comparing with Eqs.~\eqref{eq:SEss'} and \eqref{eq:Tmat}, we find that we can exactly manipulate Eq.~\eqref{eq:vartwo} into the form $\omega-E_\mathrm{X}=\Sigma_{\sigma\sigma'}(\omega)=T_{\sigma\sigma'}(\omega)/\mathcal{A}$.

The resulting co- and cross-circular spectra are shown in Fig.~\ref{fig:mobile}. The co-circular spectrum [panels (a) and (c)] features a single peak on the diagonal at $\omega_3\approx-\omega_1=E_{\rm X}$. This is qualitatively the same as the co-circular spectrum in Fig.~\ref{fig:simple}(a,c), which was calculated for immobile excitons. Closer inspection of the real part $\Re[S^{\text{1Q}}_{\up\up}]$ in Fig.~\ref{fig:mobile}(c) reveals, however, that the exciton peak is not fully dispersive; the absolute value of the real part is asymmetric across the diagonal, with the negative wing of the peak weaker than its positive wing. Since the amplitude $|S^{\text{1Q}}_{\up\up}|$ is symmetric across the peak, this reveals that the change in phase $\theta_{\up\up}$ between these two halves is less than $\pi$. For the particular exciton density shown in Fig.~\ref{fig:mobile}, the phase difference between the maximum and minimum is greater than $2\pi/3$.
Alternatively, we can observe that the phase $\theta_{\up\up}$ at the centre of the resonance is $\pi/2$ (i.e., purely imaginary) for a perfectly dispersive peak like that in Fig.~\ref{fig:simple}, whereas the co-circular peak for mobile excitons has a central phase noticeably less than $\pi/2$ for the density chosen in Fig.~\ref{fig:mobile}. This is a signature of the excitation-induced dephasing already mentioned above in our discussion of Eq.~\eqref{eq:mobileselfenergy}; the exciton-exciton interaction serves not only to shift the resonance, but also to increase its broadening, making the lineshape of the resulting peak imperfectly dispersive. The inclusion of a scattering continuum gives rise to this dephasing, realized mathematically as an imaginary part in the self-energy, without requiring any new phenomenological fitting parameters. Our microscopic modelling is also consistent with experimental observations in, e.g., Ref.~\cite{smallwood2025}, where the appearance of excitation-induced dephasing in InGaAs double quantum wells was related to the proximity of exciton resonances to scattering continua.

\begin{figure}
\centering
\includegraphics[width=\columnwidth]{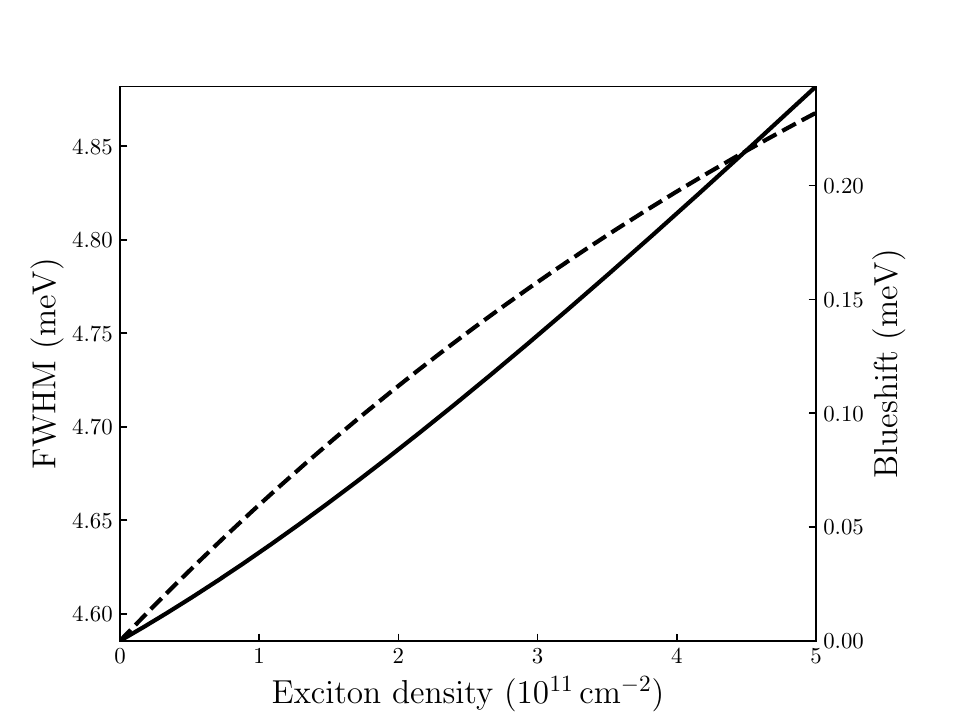}
\caption{Behavior of the unbound exciton peak in the cross-circular spectrum of mobile excitons. Full width at half maximum (solid line) and energy shift (dashed) as a function of exciton density, with homogeneous broadening $\Gamma/\varepsilon_{\rm{BX}} = 0.1$ and zero inhomogeneous broadening. Axis labels are converted to experimental units assuming a biexciton binding energy $\varepsilon_{\rm{BX}}=20$ meV.} 
\label{fig:twoproxy}
\end{figure}

Turning to the cross-circular spectrum for mobile excitons, we see that we have two peaks in Fig.~\ref{fig:mobile}(b): one on the diagonal at $\omega_3 \approx \omega_1 = E_{\rm X}$ corresponding to an unbound exciton, and one off-diagonal peak at $\omega_1 = E_{\rm X}$, $\omega_3 \approx E_{\rm{BX}} \equiv E_{\rm X}-\varepsilon_{\rm{BX}}$ corresponding to emission from the biexciton. While this resembles the cross-circular spectrum for immobile excitons in Fig.~\ref{fig:simple}(b), we emphasize that the biexciton here emerges from the scattering of two excitons rather than having to be inserted by hand as a separate state. Furthermore, differences once again appear in $\Re[S^{\text{1Q}}_{\up\up}]$, as shown in Fig.~\ref{fig:mobile}(d). Here the biexciton peak is a simple Lorentzian like before, but the exciton peak becomes imperfectly dispersive, exhibiting the same excitation-induced dephasing described above for the co-circular spectrum. 

We can analyze this further by considering the behavior as a function of exciton density. 
For the unbound exciton peak, both the energy shift and FWHM increase approximately linearly with increasing exciton density in the low-density limit, as shown in Fig.~\ref{fig:twoproxy}. Notably, this linear increase in FWHM has been observed experimentally in quantum wells~\cite{Honold1989} and in TMD monolayers~\cite{Moody2015}. It arises here as a signature of the complex self-energy which emerges from our microscopic model after the inclusion of a scattering continuum. Note that this differs from the immobile-exciton case considered above in Sec.~\ref{subsec:immobile}, for which the FWHM scales quadratically rather than linearly with the blueshift, unless an imaginary part is included \emph{ad hoc} in the exciton-exciton interaction.

By contrast, the biexciton peak redshifts with increasing exciton density, and gains spectral weight from the unbound peak. The decreasing weight of the two-exciton contribution to the unbound peak leads to a reduced cancellation with the one-exciton contribution and thus a stronger resultant peak, such that the biexciton peak and the unbound peak both strengthen in constant proportion. The same basic behavior was observed for increased coupling in the immobile model considered above. 
We note that the intensity of the biexciton peak is proportional to the (squared) zero-momentum component $\lvert \alpha_0 \rvert^2$ of the bound state, which depends in turn on the exciton density $1/\mathcal{A}$ according to $\lvert \alpha_0 \rvert^2 = 1/(1+\mathcal{A} \cdot m_{\rm X} \varepsilon_{\rm{BX}}/4\pi)$, as can be obtained from Eq.~\eqref{eq:vartwo}. At low density this gives $\lvert \alpha_0 \rvert^2 \simeq \frac{4\pi}{m_{\rm X} \varepsilon_{\rm{BX}}} \frac{1}{\mathcal{A}}$. The size of the redshift likewise scales with the density; an expansion of $-\varepsilon_{\rm{BX}}-\Delta\omega_3=\Sigma_{\uparrow\downarrow}(E_{\rm X}-\varepsilon_{\rm{BX}}-\Delta\omega_3)$ at low density yields a redshift $\Delta\omega_3\simeq\frac{1}{2}\left(\sqrt{\varepsilon_{\rm{BX}}^2 + \frac{1}{\mathcal{A}} \frac{16\pi}{m_{\rm X}} \varepsilon_{\rm{BX}}} - \varepsilon_{\rm{BX}}\right)\simeq\frac{4\pi}{m_{\rm X}}\frac{1}{\mathcal{A}}$. Thus the oscillator strength $\lvert \alpha_0 \rvert^2$ and the redshift $\Delta\omega_3/\varepsilon_{\rm{BX}}$ are equal for low densities, just as we found for the immobile model above in Sec.~\ref{subsec:immobile}. Here, however, the strength of the effect is controlled entirely by the exciton density rather than an additional microscopic parameter.

To summarize, we have demonstrated in explicit detail how a many-body Hamiltonian gives rise to a 2D spectrum, and we have observed that features of the rephasing spectrum can be directly related to features of this Hamiltonian, specifically those terms describing exciton-exciton interactions. By contrasting two models describing immobile and mobile excitons, respectively, we have observed that the introduction of continuous degrees of freedom leads to a phenomenologically richer model, with predictions concerning peak broadening, placement, and intensity emerging naturally from the microscopic model rather than being introduced by hand.

\section{Doped semiconductors} \label{sec:doped}

We now present a microscopic model of multidimensional spectroscopy on a charge-doped monolayer semiconductor. When the number of charge carriers exceeds that of the optically injected excitons, the spectrum at low temperature features new attractive and repulsive polaron quasiparticles~\cite{Sidler2017,Efimkin2017}, where the attractive polaron is adiabatically connected to the  trion (exciton--charge-carrier) bound state, while the repulsive polaron evolves into the exciton at vanishing charge doping. Multidimensional spectroscopy has already been used as a sensitive tool to measure the quasiparticle properties of single polarons in MoSe$_2$~\cite{Huang2023}, as well as to derive effective selection rules for polaron-polaron interactions in the tungsten-based monolayers WS$_2$ \cite{Muir2022} and WSe$_2$~\cite{Ni2025}, where the inverted conduction bands lead to the existence of multiple attractive polarons.

Here we focus on exciton polarons in a doped MoSe$_2$ monolayer. Specifically, we extend the model of mobile excitons examined in Sec.~\ref{subsec:mobile} by introducing a Fermi sea of itinerant charge carriers (which, for concreteness, we will henceforth assume to be electrons rather than holes). In MoSe$_2$, this Fermi sea exists both in the same valley where an exciton is created, as well as in the opposite valley. The doping in the opposite valley leads to the formation of attractive and repulsive polarons, due to the possibility of opposite-valley electrons forming trion bound states with the exciton. Conversely, because of Pauli blocking, the same-valley electron does not lead to trion formation. Hence same-valley doping primarily leads to bandgap renormalization, where the attractive and repulsive polarons shift together, as confirmed experimentally in Refs.~\cite{Sidler2017,Huang2023}. To focus on the quantum dynamics of attractive and repulsive polarons and their interplay in multidimensional spectroscopy, we therefore restrict our attention to the interaction between excitons and opposite-valley electrons. Furthermore, we will focus on electron-mediated interactions between same-spin excitons, where the competition for electron-dressing leads to a ``phase-space filling'' type of quasiparticle interaction experimentally observed in Ref.~\cite{Tan2020}. As noted above, same-spin exciton-exciton interactions are repulsive owing to Pauli exclusion effects, and do not result in a biexciton~\cite{hao2017neutral}. We therefore neglect direct exciton-exciton repulsion in our treatment here, though it is easily reintroduced at the mean-field level.

With these considerations in mind, our Hamiltonian for excitons and electrons takes the form 
\begin{align} \notag
    \ham &= \sum_{\k\sigma}(E_{\rm X} + \epsilon^{\rm X}_\k) \hat{X}^\dag_{\k\sigma} \hat{X}_{\k\sigma}^\pdag + \sum_{\k\sigma}\epsilon^{\rm f}_\k \hat f^\dag_{\k\sigma} \hat f_{\k\sigma}^\pdag \\
    &\phantom{=}~ + \frac{1}{2\mathcal{A}} \sum_{\k\k'\q\sigma} V(\q) \hat{X}^\dag_{\k+\q\sigma} \hat{f}^\dag_{\k'-\q'\bar\sigma}  \hat{f}_{\k'\bar\sigma}^\pdag \hat{X}_{\k\sigma}^\pdag \, , \label{eq:Hdope}
\end{align}
where $\hat f^\dag_{\k\sigma}$ creates a spin-$\sigma$ electron at momentum $\k$, $\epsilon^{\rm f}_{\k}=k^2/2m_{\rm f}$ is the electron dispersion (measured from the electronic bandgap), and $V(\q)$ denotes the short-ranged attractive interaction potential between excitons and electrons in opposite valleys, with $\bar\sigma \neq \sigma$. 

Following the approach presented above in Sec.~\ref{subsec:mobile}, the exciton-electron scattering $T$ matrix is obtained from the Lippmann-Schwinger equation, similarly to Eq.~\eqref{eq:Tmat}, and it takes the universal low-energy form: 
\beq\label{eq:Tmatxeuniversal}
T(\omega) = \frac{2\pi}{m_{\rm r}}\frac1{\ln\frac{\varepsilon_{\rm T}}{\omega-E_{\rm X}}+i\pi}\,,
\eeq
with reduced mass $m_{\rm r} = (1/m_{\rm X} +1/m_{\rm f})^{-1}$.
This $T$ matrix diverges when the collision energy $\omega$ matches the energy of the trion bound state, $E_\mathrm{X}-\varepsilon_\mathrm{T}$, where $\varepsilon_\mathrm{T}$ is the trion binding energy. In fact, we can obtain this low-energy behavior by taking the scattering potential to be constant, i.e., $V(\k)=g$, up to an ultraviolet cutoff $\Lambda$. In this case, the Lippmann-Schwinger equation satisfied by the $T$ matrix becomes
\beq \label{eq:Tmatxe}
T(\omega) = g + \frac{g}{\mathcal{A}}\sum_{\k}^\Lambda \frac{1}{\omega-E_{\rm X}-\epsilon_\k^\mathrm{X}-\epsilon_\k^\mathrm{f}} T(\omega) \, .
\eeq
Importantly, for this model of short-range interactions, the $T$ matrix is independent of the relative exciton-electron momentum, which we therefore suppress. Formally, Eq.~\eqref{eq:Tmatxe} can be inverted to find
\begin{align}\label{eq:Tmatxe2}
    T(\omega)=\left(\frac1g-\frac{1}{\mathcal{A}}\sum_\k^\Lambda \frac{1}{\omega-E_{\rm X}-\epsilon^{\rm X}_\k - \epsilon^{\rm f}_\k}\right)^{-1}\,.
\end{align}
Using the fact that the $T$ matrix diverges at the trion energy, i.e., when $\omega=E_\mathrm{X}-\varepsilon_{T}$, then allows us to relate the attractive coupling constant $g$ to the physical parameter $\varepsilon_\mathrm{T}$, which yields
\begin{align}\label{eq:1overg}
    \frac1g=-\frac{1}{\mathcal{A}}\sum_\k^\Lambda \frac{1}{\varepsilon_\mathrm{T}+\epsilon^{\rm X}_\k + \epsilon^{\rm f}_\k}\,.
\end{align}
While the momentum sums in Eqs.~\eqref{eq:Tmatxe2} and \eqref{eq:1overg} are both formally divergent when we take $\Lambda \to \infty$, the diverging terms cancel once we insert the latter into the former, allowing us to recover the universal form in Eq.~\eqref{eq:Tmatxeuniversal}. We stress that this approach to exciton-electron scattering is convenient, since it does not depend on the precise details of the interaction potential, only on the universal scattering behavior at low collision energies. In the following, whenever $g$ is introduced it should be understood as satisfying the definition in Eq.~\eqref{eq:1overg}, where we consider sufficiently large $\Lambda$ that our results are independent of the cutoff.

\subsection{Linear response of exciton polarons} \label{subsec:dop1X}

The main challenge confronting any microscopic approach to MDCS is the prohibitively large phase space of the doubly excited system; in this work, we mitigate this by adopting a variational ansatz with relatively few parameters. Specifically, when the exciton is introduced into the charged medium, it creates particle-hole excitations, and we will limit our attention to at most a single such excitation. Due to a suppression of terms with multiple fermionic excitations~\cite{Combescot2008}, this type of ansatz has been shown to provide an extremely good description of Fermi polarons (quantum impurities dressed by excitations of a Fermi sea) in many different scenarios~\cite{Chevy2006,Massignan2014,Scazza2022,Massignan2026}, including exciton polarons~\cite{Sidler2017,Efimkin2017,Efimkin2021,Huang2023}. Compared with these previous works, we will make the further approximation that the dynamics of the electron-hole excitation is less important than the dynamics of the particle excitation (conceptually similar treatments of hole dynamics can be found in Refs.~\cite{Suris2001} and~\cite{Glazov2020}). This approximation is strictly valid when the Fermi energy $E_{\rm F}$ is much smaller than the trion binding energy $\varepsilon_{\rm T}$. However, as we now demonstrate, our approach proves accurate even when $E_{\rm F}\sim \varepsilon_\mathrm{T}.$

Hence, to capture the effects of the coherent dressing of excitons by a Fermi sea of dopant electrons in the opposite valley, we adopt the following ansatz for a spin-$\up$ exciton polaron:
\beq 
    \ket{\psi} = \alpha_0 \hat{X}^\dag_{0\up} \FS + \frac{1}{\sqrt{\mathcal{A}N}} \sum_{\k\q} \alpha_{\k} \hat{X}^\dag_{\k+\q\up}\hat{f}^\dag_{-\k\down}\hat{f}_{\q\down}\FS . \label{eq:ans1X}
\eeq
Here $\FS = \prod_{\lvert \q \rvert < k_{\rm F}} \hat{f}^\dag_{\q\down} \ket{0}$ is a Fermi sea, where the Fermi momentum $k_{\rm F}$ is related to the number $N$ of spin-$\down$ electrons within the area $\mathcal{A}$ via the density $n=N/\mathcal{A}=k_{\rm F}^2/4\pi$. We will henceforth assume that momenta labelled $\q$ lie below the Fermi surface and momenta labelled $\k$ lie above it, i.e., $|\q|\leq k_{\rm F}< |\k|$. 
By taking $\partial_{\alpha^*} \bra{\psi} (\ham - E) \ket{\psi} = 0$, we obtain equations of motion for the exciton polaron:
\begin{subequations} \label{eq:var1X}
\begin{align}
    (E - E_{\rm X})\alpha_0 &= gn \alpha_0 + \frac{g\sqrt{n}}{\mathcal{A}} \sum_{\k'}^\Lambda \alpha_{\k'} \,,\label{eq:var1X1}\\
    (E-E_\k - E_{\rm X}) \alpha_\k &= g\sqrt{n} \alpha_0 + \frac{g}{\mathcal{A}} \sum_{\k'}^\Lambda \alpha_{\k'}
    \label{eq:var1X2}
    \,.
\end{align}
\end{subequations}
Here we have defined $E_\k = \epsilon^{\rm X}_\k + \epsilon^{\rm f}_\k + \pi n(m_{\rm X}^{-1}-m_{\rm f}^{-1})$.
Converting sums to integrals and manipulating these equations, we find that an exciton injected into the Fermi sea at energy $E=\omega$ satisfies 
the equation $(\omega - E_{\rm X})\alpha_0 = \Sigma(\omega) \alpha_0$, with the self-energy of the dressed exciton:
\beq
    \Sigma(\omega) = \frac{E_{\rm F} (1+r)}{ \ln\left[-\varepsilon_{\rm T}/(\omega - E_{\rm X} - E_{\rm F}(\frac{3}{2}r + \frac{1}{2}))\right]}. \label{eq:qSelf}
\eeq
Here $r=m_{\rm f}/m_{\rm X}$ is the mass ratio, and $E_\mathrm{F}=k_\mathrm{F}^2/2m_\mathrm{f}$ is the Fermi energy. In the following, we assume equal electron and hole masses, such that $r = 1/2$, since this is approximately the case in TMD monolayers. 

\begin{figure}
\centering
\includegraphics[width=0.85\columnwidth]{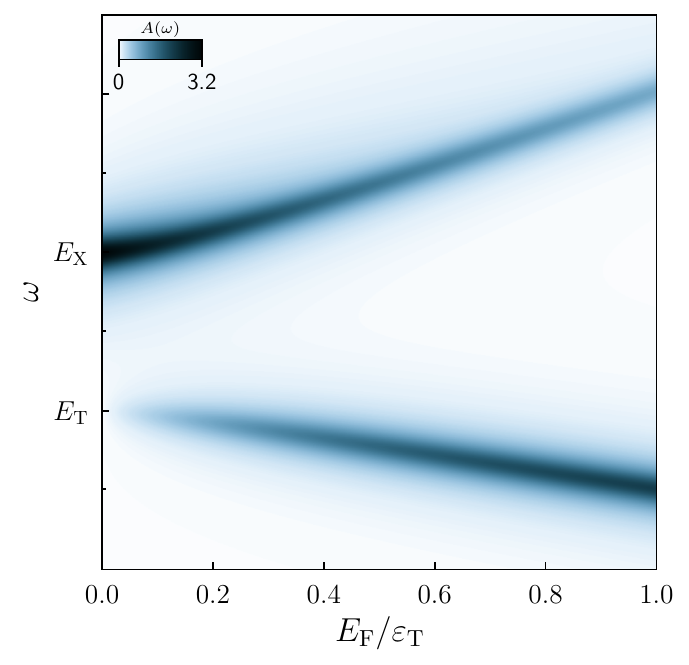}
\caption{Spectral function  in Eq.~\eqref{eq:specfunc} obtained from the ansatz in Eq.~\eqref{eq:ans1X} as a function of Fermi energy $E_{\rm F}$. A small broadening $\Gamma/\varepsilon_{\rm T} = 0.1$ has been added for visualization. As the Fermi energy increases, spectral weight shifts from the repulsive polaron to the attractive polaron.}
\label{fig:split}
\end{figure}

\begin{figure}
\centering
\includegraphics[width=\columnwidth]{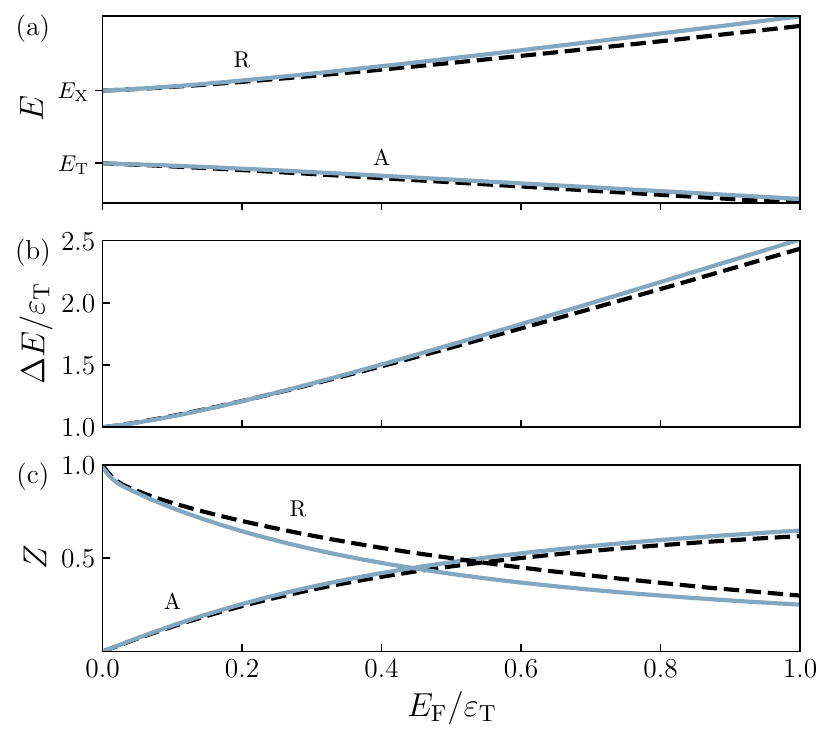}
\caption{Quasiparticle properties of attractive (A) and repulsive (R) exciton polarons: (a) polaron energy versus Fermi energy, (b) the polaron energy splitting $\Delta E=E_\mathrm{R}-E_\mathrm{A}$, and (c) the corresponding quasiparticle residues $Z=|\alpha_0|^2$. Solid lines correspond to our variational ansatz in Eq.~\eqref{eq:ans1X}, which neglects the dynamics associated with the Fermi-sea hole, while the dashed line is the Chevy ansatz~\cite{Chevy2006} which includes these dynamics (see Appendix~\ref{app:chevy}).}
\label{fig:1Xprops}
\end{figure}

The linear absorption spectrum is proportional to the spectral function 
\begin{align}\label{eq:specfunc}
    A_\up(\omega) = -\frac1\pi\Im\left[ G_\up(\omega) \right]=-\frac1\pi \Im\frac1{\omega-E_\mathrm{X}-\Sigma(\omega)}\,.
\end{align}
As shown in Fig.~\ref{fig:split}, the spectrum features two clear peaks corresponding to (real) solutions of $\omega-E_\mathrm{X}=\Sigma(\omega)$, one with peak energy $E_{\rm A} \le E_{\rm T}\equiv E_\mathrm{X}-\varepsilon_\mathrm{T}$ (the attractive polaron, continuously connected to the trion), and one with energy $E_{\rm R} \ge E_{\rm X}$ (the repulsive polaron, connected to the exciton).

It is instructive to compare the results of our ansatz with those obtained using variational parameters that are sensitive to hole momenta, i.e., taking $\alpha_{\k}\to\alpha_{\k\q}$ in Eq.~\eqref{eq:ans1X}. Such an ansatz was first introduced in Ref.~\cite{Chevy2006} to describe atomic polarons, and while this ansatz is by now fairly standard, for completeness we include the relevant equations in Appendix~\ref{app:chevy}.
As illustrated in Fig.~\ref{fig:1Xprops}, there is close agreement between the two ans\"{a}tze for the quasiparticle properties, even up to $E_\mathrm{F}\sim \varepsilon_\mathrm{T}$. Specifically, we show the energies of the attractive and repulsive branches, their energy splitting, and their quasiparticle residues (overlap $Z=|\alpha_0|^2$ with the ground state of the non-interacting system) which characterize the spectral weights of the two branches. 
Importantly, our model reproduces the characteristic approximately linear splitting $E_{\rm R} - E_{\rm A} \approx \varepsilon_{\rm T} + \frac{3}{2} E_{\rm F}$~\cite{Efimkin2021,Huang2023}. 

The most notable difference between the two ans\"atze is that our upper branch at $E_{\rm R}$ is an exact eigenstate with an infinite lifetime, i.e., $\Im[\Sigma(E_{\rm R})]=0$, and thus it lacks the many-body dephasing which is known to characterize repulsive Fermi polarons~\cite{Adlong2020} and which appears when using the full ansatz (though note that this fails to reproduce the experimentally observed broadening in monolayer TMDs~\cite{Huang2023}). We can understand this difference as follows: By averaging over hole momenta, we exclude low-energy states wherein the hole is very close to the Fermi surface. This in turn shifts the continuum of scattering states (eigenfunctions which are sharply peaked around a particular $\alpha_\k$ in momentum space and thus approximately satisfy $\omega=E_{\rm X} + E_\k$) above the repulsive polaron. Thus, the repulsive polaron is not immersed in a continuum of scattering states and does not undergo dephasing in this approximation. We also find that the quasiparticle residue of the repulsive polaron, $Z_+$, is less than $1-Z_-$, with $Z_-$ the attractive polaron residue, which indicates that the remaining residue, $1-(Z_++Z_-)$, is distributed over the incoherent continuum of scattering states. This feature is also present in the Chevy ansatz, although in that case slightly less spectral weight is transferred from the repulsive polaron to the continuum.

To summarize, having omitted hole momentum $\q$ in our variational parameters in Eq.~\eqref{eq:ans1X}, we search a considerably lower-dimensional variational subspace without seriously compromising the physical fidelity of our result within the linear response regime. As we will see below, the use of ans\"{a}tze similar to Eq.~\eqref{eq:ans1X} allows us to model the two-polaron system and to reproduce phenomenological features beyond the reach of typical few-level models.

\subsection{Multi-dimensional spectroscopy} \label{subsec:dop2X}
We now seek to model a co-circularly polarized MDCS experiment in a doped MoSe$_2$ monolayer. As emphasized in Section~\ref{sec:simple}, any nonzero 1Q rephasing signal is a signature of interactions between two exciton polarons. We limit our attention to the interaction due to phase-space filling in the Fermi sea---if an electron is drawn into the dressing cloud of one exciton, it is thereby excluded from a second exciton's dressing cloud, in effect reducing the electron density experienced by the second exciton. 

\begin{figure*}
\centering
\includegraphics[width=0.85\textwidth]{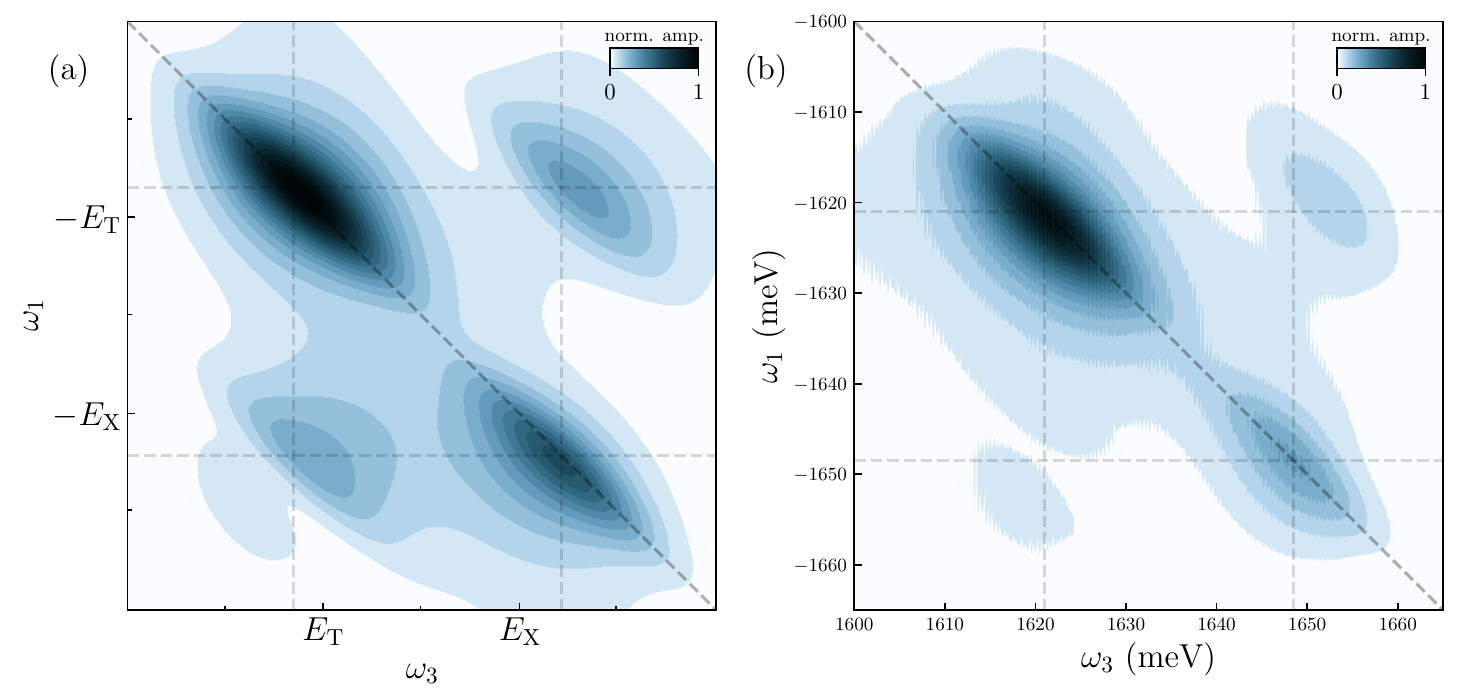}
\caption{Absolute value of co-circular 1Q rephasing spectrum with time delay $t_2=0$. (a) Microscopic model with Fermi energy $E_{\rm F}/\varepsilon_{\rm T}\approx0.327$, homogeneous broadening $\Gamma/\varepsilon_{\rm T}=0.12$, and inhomogeneous broadening $\gamma/\varepsilon_{\rm T}=0.25$. (b) Experimental spectrum for doped MoSe$_2$, data from Ref.~\cite{hao2017neutral}.}
\label{fig:compSpec}
\end{figure*}

To describe the doubly excited system, we extend the single-exciton ansatz above: 
\begin{widetext}
\begin{align}
    \ket{\psi} &= \frac{1}{\sqrt{2}} \alpha_0 \hat{X}^\dag_{0\up} \hat{X}^\dag_{0\up} \FS + \frac{1}{\sqrt{\mathcal{A}N}} \sum_{\k\q} \alpha_\k \hat{X}^\dag_{0\up} \hat{X}^\dag_{\k+\q\up} \hat{f}^\dag_{-\k\down} \hat{f}^\pdag_{\q\down} \FS \notag\\
    &~+ \frac{1}{2\mathcal{A}\sqrt{N(N-1)}} \sum_{\k_1 \k_2 \q_1 \q_2} \alpha_{\k_1\k_2} \hat{X}^\dag_{\k_1+\q_1\up} \hat{X}^\dag_{\k_2 + \q_2\up} \hat{f}^\dag_{-\k_1\down} \hat{f}^\pdag_{\q_1\down} \hat{f}^\dag_{-\k_2\down} \hat{f}^\pdag_{\q_2\down} \FS\,, \label{eq:2Xans}
\end{align}
\end{widetext}
where we require $\alpha_{\k_1\k_2} = \alpha_{\k_2\k_1}$ due to Bose statistics. In keeping with our general emphasis on the effects of resonantly enhanced two-body physics (associated with the trion bound state), the ansatz contains only those terms involving pairwise scattering processes. Importantly, the two excitons are dressed by electron-hole pairs from the same Fermi sea, enabling the phase-space filling effect described above. The action of a particle-hole operator on the Fermi sea $\FS$ removes a particle from below the Fermi surface, and consequently the coupling between the $\alpha_\k$ and $\alpha_{\k_1\k_2}$ terms in our variational ansatz is proportional not to the electron density $n$ but to the depleted electron density $n_{-}\equiv (N-1)/\mathcal{A}$. We will neglect electron-electron exchange and exciton-exciton exchange interactions, as done in Ref.~\cite{Tan2020}, since these approximately cancel one another for low Fermi energy $E_{\rm F} \lesssim \varepsilon_{\rm T}$. 

Using the ansatz in Eq.~\eqref{eq:2Xans}, we arrive at the following equations of motion:
\begin{subequations} \label{eq:var2X}
\begin{align}
    &(E - 2E_{\rm X})\alpha_0 = 2gn \alpha_0 + \frac{g\sqrt{2n}}{\mathcal{A}}\sum_{\k'} \alpha_{\k'}\,, \label{eq:var2X1}\\ \notag
    &(E-E_\k - 2E_{\rm X}) \alpha_\k = g \sqrt{2n} \alpha_0 + gn \alpha_\k + \frac{g}{\mathcal{A}}\sum_{\k'} \alpha_{\k'}\\
    &\phantom{E-E_\k - 2E_{\rm X}) \alpha_\k =} + \frac{g\sqrt{n_{-}}}{\mathcal{A}} \sum_{\k'} \alpha_{\k\k'}\,, \label{eq:var2X2}\\ \notag
    &(E-E_{\k_1}-E_{\k_2} - 2E_{\rm X}) \alpha_{\k_1 \k_2} = g\sqrt{n_{-}} (\alpha_{\k_1} + \alpha_{\k_2}) \\
    &\phantom{(E-E_{\k_1}-E_{\k_2} - 2E_{\rm X}) \alpha_{\k_1 \k_2} = } + \frac{g}{\mathcal{A}} \sum_{\k'} (\alpha_{\k_1 \k'} + \alpha_{\k' \k_2})\,. \label{eq:var2X3}
\end{align}
\end{subequations}
Within our model, the appearance of the depleted electron density $n_{-}$ in Eqs.~\eqref{eq:var2X2} and~\eqref{eq:var2X3} is the sole source of coupling between the two excitons. If we took the number of available electrons to be constant and set $n_{-}=n$ instead, our solutions would factor into symmetrized products of single-exciton solutions (i.e., solutions to Eqs.~\eqref{eq:var1X1}-\eqref{eq:var1X2}) and the 1Q rephasing signal would necessarily vanish. Thus, every feature of our simulated spectra will be a consequence of phase-space filling in the Fermi sea.

\begin{figure}
\centering
\includegraphics[width=0.9\columnwidth]{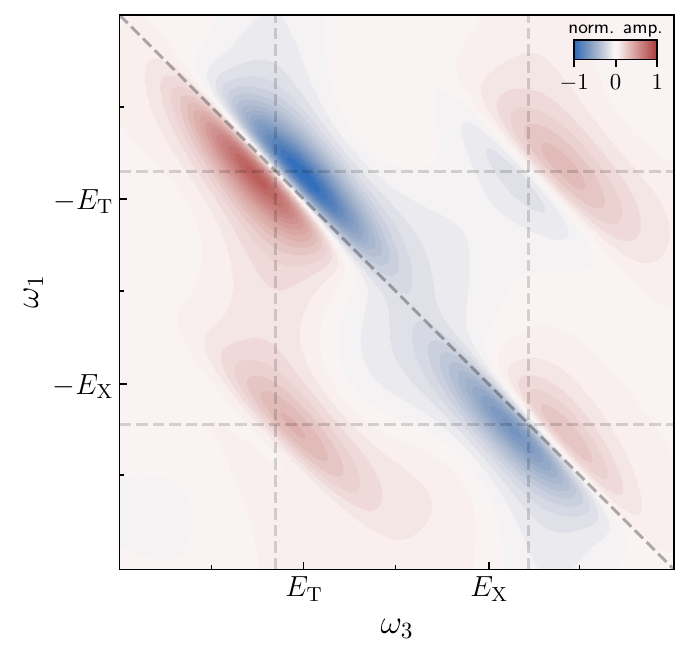}
\caption{Real part of co-circular 1Q rephasing spectrum shown in  Fig.~\ref{fig:compSpec}(a). Positive peaks (red) show contributions from ground state bleach and stimulated emission, while negative peaks (blue) show contributions from excited-state absorption.}
\label{fig:spec50re}
\end{figure}

\begin{figure*}
\centering
\includegraphics[width=0.8\textwidth]{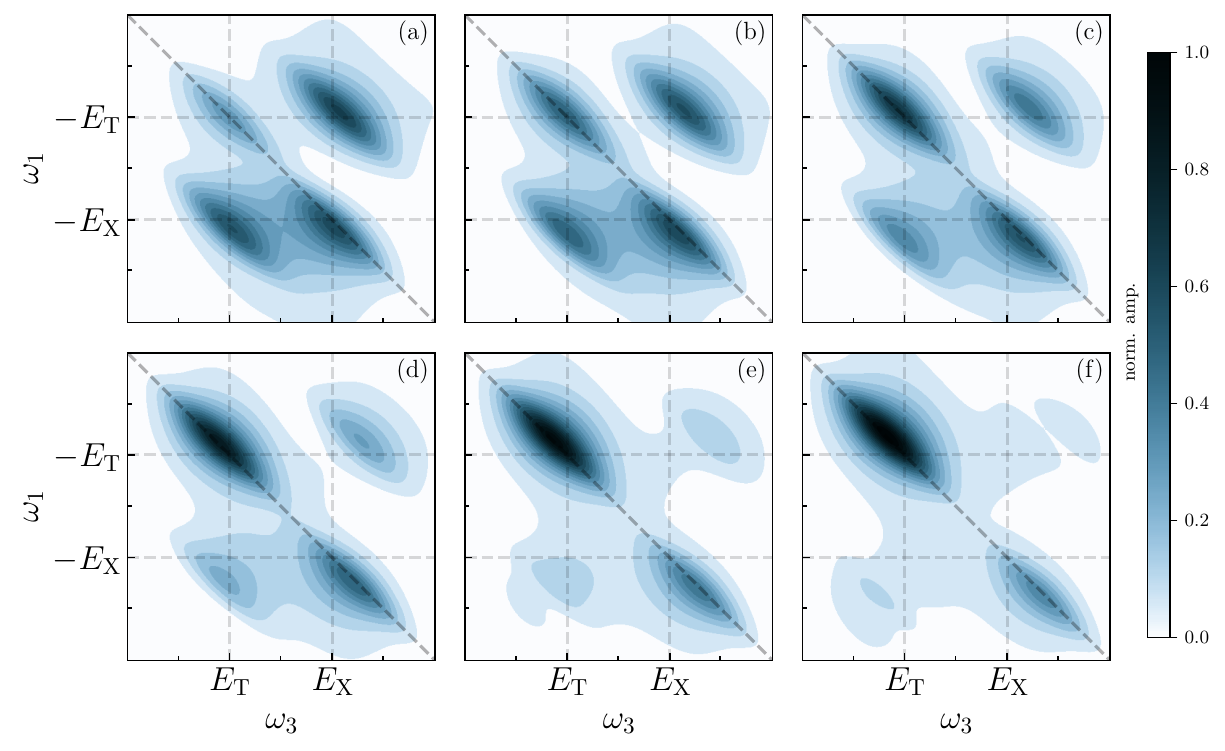}
\caption{Absolute value of 1Q rephasing spectrum for a range of doping values, with Fermi energy $E_{\rm F}/\varepsilon_{\rm T} \approx 0.13$, $0.19$, $0.25$, $0.31$, $0.38$, $0.44$, respectively, for (a)-(f). All spectra share a common normalization such that their amplitudes are directly comparable.}
\label{fig:dope}
\end{figure*}

We solve the variational equations in Eq.~\eqref{eq:var2X} numerically on a discrete momentum grid, and use the results to evaluate the third-order response functions in Sec.~\ref{sec:theory}. As an example, the absolute value of a calculated 1Q rephasing spectrum is shown in Fig.~\ref{fig:compSpec} (corresponding 0Q and 2Q spectra are presented in Appendix~\ref{app:morespectra}), alongside experimental data from Ref.~\cite{hao2017neutral}. 
Here we have used parameters corresponding to an exciton density of $1.3\times10^{9}\,\text{cm}^{-2}$ and an electron density of $3.45 \times 10^{10}\,\text{cm}^{-2}$, and assumed a trion binding energy $\varepsilon_{\rm T} = 25\text{ meV}$. These parameters, which are well within the typical experimental range, were chosen to achieve a reasonable fit with the experiment for which precise density values were not available. Overall, we observe a strong agreement between the simulated spectrum and experimental data. Qualitatively, the expected diagonal peaks corresponding to the repulsive and attractive polarons are observed, as well as cross-peaks indicating coherent coupling between the two branches. Quantitatively, the relative peak intensities are reproduced remarkably well.

At this moderate level of doping, with Fermi energy $E_{\rm F}/\varepsilon_{\rm T}= 0.327$, a hierarchy of peak intensities is seen: The diagonal attractive polaron (AP) peak is the strongest, followed by the diagonal repulsive polaron (RP) peak, followed by the two cross-peaks. This matches the hierarchy observed experimentally for doped MoSe$_2$ in Ref.~\cite{hao2017neutral}. As we have noted above in Sec.~\ref{sec:simple}, the intensity of each peak is dependent not only on the oscillator strength of the states but also on the strength of their interactions.
These relative intensities are strongly dependent on the Fermi energy; a different hierarchy holds for lower doping where $E_{\rm F}\lesssim0.25\varepsilon_{\rm T}$.

We also note an intriguing asymmetry and fine structure in the two cross-peaks. This is a consequence of constructive and destructive interference between the quasiparticle cross-peaks, positioned in the lower-left at $(E_{\rm A},E_{\rm R})$ and in the upper-right at $(E_{\rm R},E_{\rm A})$, and weaker contributions from the tail of high-energy continuum states above the repulsive polaron, displaced below the lower-left peak and to the right of the upper-right peak. This is best understood by examining the real part of the theoretical spectrum, shown in Fig.~\ref{fig:spec50re}. The four main peaks exhibit a `dispersive' profile with a phase shift of $\pi$ (i.e., a sign change) along the $\omega_3$ axis, as is typical of peaks resulting from excitation-induced shifts in the resonance. At the doping level shown in Fig.~\ref{fig:spec50re}, the two off-diagonal peaks change from negative to positive as $\omega_3$ increases and thus result from attractive AP-RP interactions. On the other hand, the continuum state contributions are predominantly positive, so they interfere constructively with the upper-right peak, leading to a stronger peak with a lengthened tail in the direction of increasing $\omega_3$, and destructively with the lower-left peak, leading to a weaker peak with a shortened tail in the direction of decreasing $\omega_1$. This explains the difference in cross-peak shapes and intensities. 

It is interesting to note that cross-peak asymmetry is observed both in the results of Ref.~\cite{hao2017neutral}, shown in Fig.~\ref{fig:compSpec}(b), and in the results of Ref.~\cite{Hao2016}. In both experiments, the upper-right cross-peak was found to be stronger than the lower-left cross-peak, consistent with our prediction of constructive interference with continuum states in the upper right and destructive interference in the lower left. We also note that the upper-right cross-peak in Fig.~\ref{fig:compSpec}(b) is somewhat blueshifted in emission frequency compared to the lower-right diagonal RP peak, a feature which our calculated spectrum reproduces by virtue of constructive interference with continuum states. There are also indications that the lower-left cross-peak in Fig.~\ref{fig:compSpec}(b) exhibits a central node, as we would expect from the split-peak structure caused by destructive interference with the continuum. Thus, the fine structure of the experimentally observed cross-peaks appears to agree with our description of the scattering continuum.

To gain further insight into these results, we show a range of doping regimes in Fig.~\ref{fig:dope}. As in the linear spectrum in Fig.~\ref{fig:split}, the splitting between the two branches (and thus the inter-peak distance) visibly increases with doping, and spectral weight is seen to shift from the repulsive branch to the attractive branch. Furthermore, at higher doping than that shown in Fig.~\ref{fig:compSpec}, the dispersive profiles of the cross-peaks change sign, indicating that the AP-RP interaction changes from attractive to repulsive. As a consequence, the lower-left peak becomes the stronger, as clearly seen in panels (e) and (f) of Fig.~\ref{fig:dope}. This is accompanied by the appearance of a node in the upper-right peak, a signature of destructive interference with the continuum.
Interestingly, we find that the relative intensity of the cross-peaks appears to be overestimated at low doping when compared with the results in Ref.~\cite{Hao2016}. This could be explained by the lack of any direct exciton-exciton repulsion in our model, which at low doping would lead to a dominant diagonal RP peak (we see  below that the RP-RP attraction due to phase-space filling vanishes at zero doping while the AP-RP interaction strengthens).

\begin{figure} 
\centering
\includegraphics[width=\columnwidth]{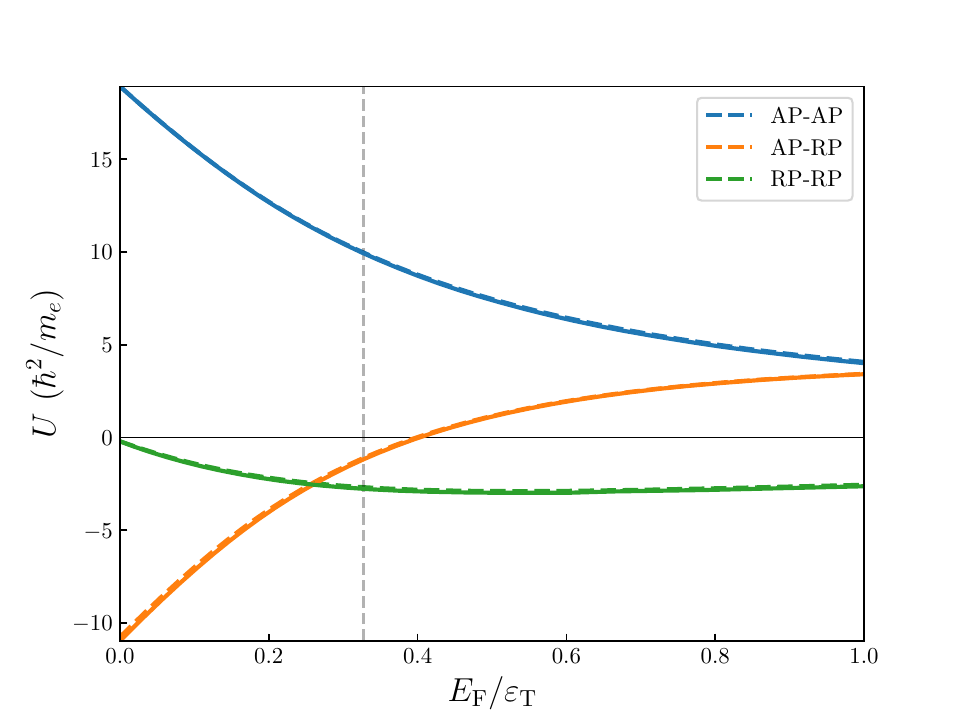}
\caption{Strengths of interactions between two attractive polarons (blue), between an attractive and a repulsive polaron (orange), and between two repulsive polarons (green). Solid lines are differences in eigenenergies between Eqs.~\eqref{eq:var2X} and~\eqref{eq:var1X}, while dashed lines are calculated from eigenstates of Eqs.~\eqref{eq:var1X} alone using Eq.~\eqref{eq:qpintborn}.
The vertical dashed line indicates the doping level used in Figs.~\ref{fig:compSpec} and~\ref{fig:spec50re}.}
\label{fig:seaEIS}
\end{figure}

We have emphasized that all spectral features are a consequence of interactions between exciton polarons. We can calculate the strength of these interactions by determining the doping-dependent differences between the solutions of the two-polaron equations of motion~\eqref{eq:var2X} and those for single polarons in Eq.~\eqref{eq:var1X}. These interaction-induced shifts in eigenenergies can be captured by the Born approximation, which corresponds to the expectation value~\cite{Levinsen2019a} 
\beq\label{eq:qpintborn}
\frac{U_{ij}}{\mathcal{A}} = \frac{\expval{\hat{a}^\pdag_i\hat{a}^\pdag_j(\ham-E_i-E_j)\hat{a}^\dag_j\hat{a}^\dag_i}}{1+\delta_{ij}}\,,
\eeq
where $\hat{a}^\dag_i = \alpha_0^{(i)}\hat{X}^\dag_{0\up} + \frac{1}{\sqrt{\mathcal{A}}}\sum_\k \alpha_\k^{(i)} \hat{X}^\dag_{\k\up} \hat{f}^\dag_{-\k\down} \hat{f}^\pdag_{\q\down}$ creates the $i$'th single-exciton eigenstate with energy $E_i$, and we have accounted for the statistical factor when the bosonic excitons are identical. The details of this calculation are shown in Appendix~\ref{app:polint}. Again, as above in Eq.~\eqref{eq:var2X}, we neglect terms arising from electron-electron exchange and exciton-exciton exchange processes. The same calculation was done using the Chevy ansatz in Ref.~\cite{Tan2020} to obtain the AP-AP interaction strength. Since the repulsive polaron is an exact eigenstate within our approximation, we can easily calculate RP-RP and AP-RP interactions as well. 

The resulting interaction-induced shifts are shown in Fig.~\ref{fig:seaEIS}. Consistent with Ref.~\cite{Tan2020}, we find a repulsive interaction between attractive polarons which weakens with increased doping. Our calculations differ from those in Ref.~\cite{Tan2020} by less than 16\% at high doping $E_{\rm F}/\varepsilon_{\rm T}=1$, a discrepancy which monotonically decreases as doping is lowered, falling below 10\% for moderate doping $E_{\rm F}/\varepsilon_{\rm T}<0.5$. For the purposes of calculating polaron-polaron interactions within the regime of interest, then, the two ans\"{a}tze are in close agreement. We also find an attractive interaction between repulsive polarons, negligible at low doping ($E_{\rm F} \sim 0$) but strengthening at intermediate doping ($E_{\rm F}/\varepsilon_{\rm T}<0.5$) before weakening again. The interaction between attractive and repulsive polarons displays a crossover from attraction at low doping to repulsion at high doping, with the sign change occurring at $E_{\rm F}/\varepsilon_{\rm T}\approx0.4$. This matches our qualitative observations above: The upper-right (lower-left) cross-peak exhibits constructive (destructive) interference with the continuum for low doping but crosses over to destructive (constructive) interference for doping $E_{\rm F}/\varepsilon_{\rm T}>0.4$, matching the change from attractive to repulsive AP-RP interactions.

\section{Conclusions and Outlook}
\label{sec:conc}
In summary, we have demonstrated the advantages of applying a microscopic theory to the multidimensional spectroscopy of 2D semiconductors. Specifically, we have shown that including the continuum associated with mobile excitons and employing a formalism based on a universal low-energy description of interactions allows us to model features that are often observed in experiments, such as peak placement, intensity, and broadening, without the need to introduce an extensive set of fitting parameters. In addition, we have shown how the presence of a scattering continuum can lead to constructive or destructive interference of cross-peaks, again a common experimental feature. Finally, our model of mediated interactions between exciton polarons in MoSe$_2$ monolayers---which incorporates resonantly enhanced electron-exciton interactions and phase-space filling in the Fermi sea---produces a phenomenologically rich 1Q rephasing spectrum in excellent agreement with experiment~\cite{hao2017neutral}.

Our results strongly suggest that microscopic models featuring continuum degrees of freedom can yield new, testable predictions for a range of semiconductor systems, such as novel ways of extracting quasiparticle interactions. For instance, extensions of our model potentially enable further insights into the interactions between singlet and triplet trions (attractive polarons) in WS$_2$ and WSe$_2$ in experiments~\cite{Muir2022,Ni2025}, the formation of larger bound complexes (and their polaronic extensions) such as charged biexcitons~\cite{hao2017neutral}, or the precise nature of the states that appear at very large doping in the tungsten-based TMDs~\cite{Jones2013}. Similarly, they provide a natural framework for analysing the interactions between Rydberg excitons in TMDs~\cite{Chernikov2014}, in cuprous oxide~\cite{Kazimierczuk2014}, or even between Rydberg exciton polarons in doped semiconductors~\cite{LiuRydberg2021}. These examples highlight the broader potential of microscopic approaches and MDCS to clarify interaction mechanisms across diverse semiconductor systems.

\acknowledgments 
We acknowledge useful discussions with Olivier Bleu, Jared Cole, Dmitry Efimkin, Josh Gray, Elaine Li and Brendan Mulkerin, and we thank Elaine Li, Galan Moody and Kai Hao for sharing the data of Ref.~\cite{hao2017neutral}.
MMP JAD and JL acknowledge support from the Australian Research Council (ARC) Centre of Excellence in Future Low-Energy Electronics Technologies (CE170100039). MMP and JL also acknowledge support from ARC Discovery Projects DP240100569 and DP250103746.
MMP is also supported through an ARC Future Fellowship FT200100619. NW acknowledges
support from an Australian Government Research Training Program (RTP) Scholarship.

\appendix 

\begin{widetext}

\section{Double quantum response for immobile excitons} \label{sec:2Qappendix}

Here we include calculations of the double quantum response for the system of immobile excitons described in Sec.~\ref{subsec:immobile}. Inserting Eq.~\eqref{eq:Hsimple} into Eq.~\eqref{eq:P3expt2Q}, we easily obtain the co-circular and cross-circular 2Q response functions in the time domain:
\begin{subequations} \label{eq:2Qimmobile}
\begin{align}
&    S^{\text{2Q}}_{\up\up}(t_3,t_2,t_1) = \, 2 e^{-iE_{\rm{X}} (t_1 +2t_2 + t_3)} e^{-i U t_2} (1 - e^{-i U t_3}) \,,\\ 
    & S^{\text{2Q}}_{\up\down}(t_3,t_2,t_1) =   e^{-iE_{\rm{X}} (t_1 +2t_2 + t_3)} [u^2 e^{-i E_+ t_2}(1-e^{-i E_+ t_3}) + v^2 e^{-i E_- t_2}(1-e^{-i E_- t_3})] \,.
\end{align}
\end{subequations}
We take a double Fourier transform to find the frequency-domain response functions:
\begin{subequations}
\begin{align}
&    S^{\text{2Q}}_{\up\up}(\omega_3,\omega_2,t_1) = \, \frac{2e^{-i(E_{\rm X}-i\Gamma) t_1}}{\omega_2 - (2E_{\rm X} + U) + i\Gamma}\left( \frac{1}{\omega_3 - E_{\rm X} + i\Gamma} - \frac{1}{\omega_3 - (E_{\rm X} + U) + i\Gamma} \right) \,,\\  \notag
    & S^{\text{2Q}}_{\up\down}(\omega_3,\omega_2,t_1) =   \frac{u^2 e^{-i(E_{\rm X}-i\Gamma) t_1}}{\omega_2 - (E_{\rm X} + E_+) + i\Gamma} \left( \frac{1}{\omega_3 - E_{\rm X} + i\Gamma} - \frac{1}{\omega_3 - (E_{\rm X} + E_+) + i\Gamma} \right) \\
   & + \frac{v^2 e^{-i(E_{\rm X}-i\Gamma) t_1}}{\omega_2 - (E_{\rm X} + E_-) + i\Gamma} \left( \frac{1}{\omega_3 - E_{\rm X} + i\Gamma} - \frac{1}{\omega_3 - (E_{\rm X} + E_-) + i\Gamma} \right) \,.
\end{align}
\end{subequations}
\end{widetext}
We see that the 2Q response will vanish both in the non-interacting limit ($U\to0$, $\lambda\to0$) and in the strongly interacting limit ($U\to\infty$, $\lambda\to\infty$).

\section{Chevy ansatz for exciton polarons}
\label{app:chevy}

Here we present the well-known Chevy ansatz as a description of exciton polarons in two dimensions, where the system of excitons and electrons is governed by the Hamiltonian Eq.~\eqref{eq:Hdope} in Sec.~\ref{sec:doped}. The Chevy ansatz differs from the ansatz Eq.~\eqref{eq:var1X} only in that it uses variational parameters $\varphi_{\k\q}$ which are sensitive to hole momenta~\cite{Chevy2006}: 
\beq 
    \ket{\psi} = \alpha_0 \hat{X}^\dag_{0\up} \FS + \frac{1}{\mathcal{A}} \sum_{\k,\q} \varphi_{\k\q} \hat{X}^\dag_{\k+\q\up}\hat{f}^\dag_{-\k\down}\hat{f}_{\q\down}\FS . \label{eq:ans1Xfull}
\eeq
We obtain the equations of motion:
\begin{subequations}
\begin{align}
    (E - E_{\rm X})\alpha_0 &= gn \alpha_0 + \frac{g}{\mathcal{A}^2} \sum_{\k'\q'}^\Lambda \varphi_{\k'\q'} \label{eq:var1X1chevy},\\
    (E-E_{\k\q} - E_{\rm X}) \alpha_{\k\q} &= g \alpha_0 + \frac{g}{\mathcal{A}} \sum_{\k'}^\Lambda \alpha_{\k'\q}
    \label{eq:var1X2chevy},
\end{align}
\end{subequations}
where $E_{\k\q}= \epsilon^{\rm X}_{\k+\q} + \epsilon^{\rm f}_{\k} - \epsilon^{\rm f}_{\q}$. We again take $E=\omega$ and rearrange these equations to the form $(\omega-E_\mathrm{X})\alpha_0=\Sigma(\omega)$, with the self-energy
\beq
    \Sigma(\omega) = \frac{1}{\mathcal{A}}\sum_{\q} \left[ \frac{1}{g} - \frac{1}{\mathcal{A}} \sum_\k \frac{1}{\omega-E_{\k\q} - E_{\rm X}} \right]^{-1}\,. \label{eq:chSelf}
\eeq
From the self-energy, we can obtain the quasiparticle properties shown in Fig.~\ref{fig:simple}. Here, we note that within the Chevy ansatz where the continuum starts at the exciton, the repulsive polaron energy is found from $E_\mathrm{R}-E_\mathrm{X}=\mathrm{Re}[\Sigma(E_\mathrm{R})]$, and the residue from
$Z_+=(1-\left.\partial_\omega\mathrm{Re}[\Sigma(\omega)]\right|_{\omega=E_{\rm R}})^{-1}$.

\section{Zero and double quantum spectra for doped monolayer semiconductors}
\label{app:morespectra}

\begin{figure}
\centering
\includegraphics[width=\columnwidth]{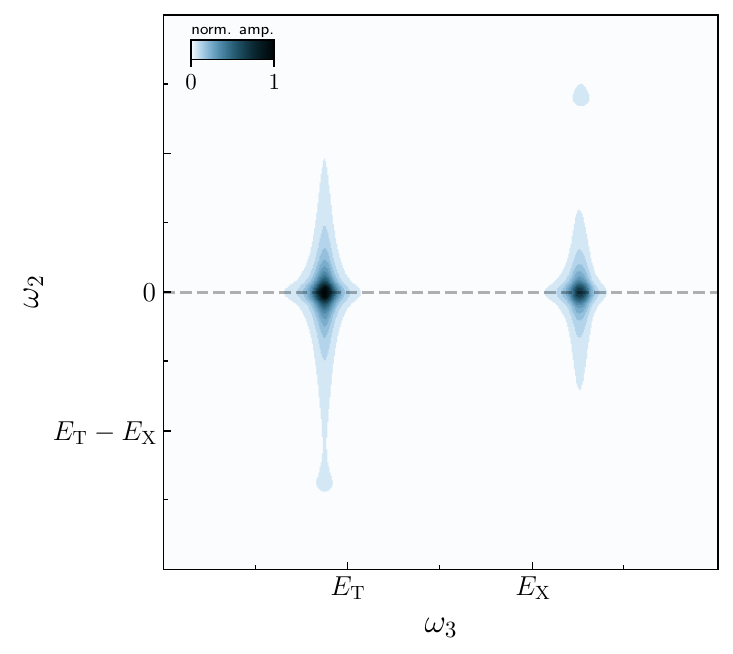}
\caption{Absolute value of co-circular 0Q rephasing spectrum with time delay $t_1=0$, Fermi energy $E_{\rm F}/\varepsilon_{\rm T} \approx 0.327 $, homogeneous broadening $\Gamma/\varepsilon_{\rm T}=0.05$, and inhomogeneous broadening $\gamma/\varepsilon_{\rm T} = 0$.}
\label{fig:0Q}
\end{figure}

\begin{figure}
\centering
\includegraphics[width=\columnwidth]{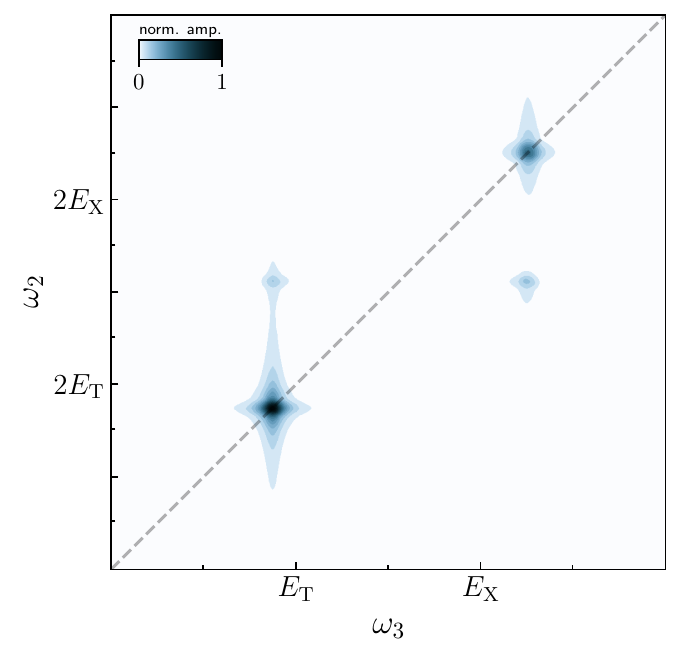}
\caption{Absolute value of co-circular 2Q non-rephasing spectrum with time delay $t_1=0$, Fermi energy $E_{\rm F}/\varepsilon_{\rm T} \approx 0.327 $, homogeneous broadening $\Gamma/\varepsilon_{\rm T}=0.05$, and inhomogeneous broadening $\gamma/\varepsilon_{\rm T} = 0$.}
\label{fig:2Q}
\end{figure}

Here we show zero quantum and double quantum spectra computed using the solutions to Eq.~\eqref{eq:var2X} in Sec.~\ref{subsec:dop2X}. Fig.~\ref{fig:0Q} shows the zero-quantum rephasing spectrum, wherein the first time delay $t_1$ is held constant and the second time delay is scanned. Processes involving a coherent superposition of different states, i.e., processes wherein the first and second pulses target different resonances, will undergo phase evolution during $t_2$ and will thus appear at nonzero $\omega_2$ corresponding to the energy difference between the two resonances. Processes involving a simple population, wherein the first two pulses have the same energy, will appear at $\omega_2=0$. 

We see that Fig.~\ref{fig:0Q} features two peaks at $\omega_2=0$, one at emission frequency $\omega_3 = E_{\rm R}>E_{\rm X}$ corresponding to a population of repulsive polarons and one at emission frequency $\omega_3 = E_{\rm A}<E_{\rm T}$ corresponding to a population of attractive polarons. We also observe two peaks indicating coherent superpositions between the two quasiparticles, one at $(\omega_3,\omega_2)=(E_{\rm A},E_{\rm A} - E_{\rm R})$ and another at $(\omega_3,\omega_2)=(E_{\rm R},E_{\rm R}-E_{\rm A})$, where we note that $E_{\rm R}-E_{\rm A}\approx E_\mathrm{X}-E_\mathrm{T}\equiv\varepsilon_\mathrm{T}$.

Likewise, Fig.~\ref{fig:2Q} shows the two-quantum non-rephasing spectrum, wherein the $t_1$ delay is held constant as in the 0Q spectrum but the conjugate pulse arrives third rather than first. The first two pulses excite a coherent superposition between the ground state and a doubly excited state; peaks along the diagonal $\omega_2=2\omega_3$ correspond to processes in which both pulses excite the same transition. Off-diagonal peaks correspond to processes in which the first two pulses excite different transitions.

As we would expect, Fig.~\ref{fig:2Q} features two peaks along $\omega_2=2\omega_3$, one at emission frequency $\omega_3=E_{\rm R}$ and one at emission frequency $\omega_3=E_{\rm A}$, corresponding to processes involving two repulsive polarons and two attractive polarons, respectively. Again, we see evidence of AP-RP interactions in the off-diagonal peaks, both appearing at $\omega_2 = E_{\rm A} + E_{\rm R}$. One of these has emission frequency $E_{\rm R}$ and the other $\omega_3 = E_{\rm A}$.

\begin{widetext}
\section{Calculation of polaron-polaron interactions}
\label{app:polint}
Here we present our calculation of polaron-polaron interactions, the results of which are shown in Fig.~\ref{fig:seaEIS}. We can approximate polaron-polaron interactions by computing the following expectation value, with $\hat{a}^\dag_i = \alpha_0^{(i)}\hat{X}^\dag_{0\up} + \frac{1}{\mathcal{A}}\sum_\k \alpha_\k^{(i)} \hat{X}^\dag_{\k\up} \hat{f}^\dag_{-\k\down} \hat{f}^\pdag_{\q\down}$ an operator creating the $i$th single-exciton eigenstate with energy $\omega_i$:
\beq
\frac{U_{ij}}{\mathcal{A}} = \frac{\expval{\hat{a}^\pdag_i\hat{a}^\pdag_j(\ham-\omega_i-\omega_j)\hat{a}^\dag_j\hat{a}^\dag_i}}{1+\delta_{ij}}\,.
\eeq
The procedure of subtracting the bare quasiparticle energies ensures that we extract the $O(1/\area)$ correction due to having two excitons within the area $\area$, and it corresponds to the Born approximation~\cite{Levinsen2019a}.

As done in the context of Fermi polarons in Ref.~\cite{Tan2020}, we compute this quantity by repeated application of Wick's theorem and neglect statistical factors due to electron-electron exchange and exciton-exciton exchange. 
Decomposing our Hamiltonian into a kinetic part $\hat{H}_0$ and an interacting part $\hat{V}$, the results are as follows:
\begin{align}
    \expval{\hat{a}^\pdag_i\hat{a}^\pdag_j\hat{a}^\dag_j\hat{a}^\dag_i} &= 1 + \delta_{ij} - \frac{1}{N} \frac{1}{\mathcal{A}^2} \sum_{\k_1\k_2} \left( \lvert \alpha^{(i)}_{\k_1} \rvert^2 \lvert \alpha^{(j)}_{\k_2} \rvert^2 + \alpha^{(i)}_{\k_1}  \alpha^{(j)}_{\k_1}  \alpha^{(i)}_{\k_2} \alpha^{(j)}_{\k_2} \right)\,,
\end{align}
\begin{align}\nn
    \expval{\hat{H}_{0}} =& \expval{\hat{a}^\pdag_i\hat{a}^\pdag_j  \sum_{\k}\left((E_{\rm X} + \epsilon^{\rm X}_\k) \hat{X}^\dag_{\k\up} \hat{X}_{\k\up}^\pdag + \epsilon^{\rm f}_\k \hat{f}^\dag_{\k\down} \hat{f}_{\k\down}^\pdag \right) \hat{a}^\dag_j\hat{a}^\dag_i}\\ \nn
    =& \frac{1}{\mathcal{A}}\sum_{\k} \left( \frac{k^2}{2m_{\rm r}} + \pi n(m_{\rm X}^{-1}-m_{\rm f}^{-1}) \right)(\lvert \alpha^{(i)}_0 \rvert^2 \lvert \alpha^{(j)}_{\k} \rvert^2 + 2\alpha^{(i)}_0 \alpha^{(i)}_{\k} \alpha^{(j)}_{\k} \alpha^{(j)}_0 + \lvert \alpha^{(i)}_{\k} \rvert^2 \lvert \alpha^{(j)}_0 \rvert^2)\\
  &+ \frac{1}{\mathcal{A}^2} \sum_{\k_1,\k_2} \left( \frac{N-1}{N} \frac{k_1^2+k_2^2}{2m_{\rm r}} + \frac{2\pi(N-1)(m_{\rm X}^{-1} - m_{\rm f}^{-1})}{\mathcal{A}}\right) (\lvert \alpha^{(i)}_{\k_1} \rvert^2 \lvert \alpha^{(j)}_{\k_2} \rvert^2 + \alpha^{(i)}_{\k_1} \alpha^{(i)}_{\k_2} \alpha^{(j)}_{\k_1} \alpha^{(j)}_{\k_2})\,,
\end{align}
\begin{align}\nn
\expval{\hat{V}} =& \expval{\hat{a}^\pdag_i\hat{a}^\pdag_j  \frac{g}{\mathcal{A}} \sum_{\k\k'\q} \hat{X}^\dag_{\k+\q\up} \hat{f}^\dag_{\k'-\q'\down} \hat{X}_{\k\up}^\pdag \hat{f}_{\k'\down}^\pdag \hat{a}^\dag_j\hat{a}^\dag_i}\\ \nn
=& 4gn\lvert \alpha^{(i)}_{0} \rvert^2 \lvert \alpha^{(j)}_{0} \rvert^2 + 4g\sqrt{n} \frac{1}{\mathcal{A}}\sum_\k ( \alpha^{(i)}_{0} \alpha^{(i)}_{\k} \lvert \alpha^{(j)}_{0} \rvert^2 + \lvert \alpha^{(i)}_{0} \rvert^2 \alpha^{(j)}_{0} \alpha^{(j)}_{\k} )\\ \nn
  & + g \frac{1}{\mathcal{A}^2}\sum_{\k_1,\k_2}(2\alpha^{(i)}_{0}\alpha^{(i)}_{\k_1}\alpha^{(j)}_{0}\alpha^{(j)}_{\k_2} + \alpha^{(i)}_{\k_1}\alpha^{(i)}_{\k_2}\lvert \alpha^{(j)}_{0} \rvert^2 + \lvert \alpha^{(i)}_{0} \rvert^2\alpha^{(j)}_{\k_1}\alpha^{(j)}_{\k_2})
  \\ \nn
  &+ gn \frac{1}{\mathcal{A}}\sum_{\k}(2\alpha^{(i)}_{0}\alpha^{(i)}_{\k}\alpha^{(j)}_{0}\alpha^{(j)}_{\k} + \lvert \alpha^{(i)}_{\k} \rvert^2\lvert \alpha^{(j)}_{0} \rvert^2 + \lvert \alpha^{(i)}_{0} \rvert^2\lvert \alpha^{(j)}_{\k} \rvert^2 ) \\ \nn
  &+ 2g\frac{N-1}{\sqrt{\mathcal{A}N}} \frac{1}{\mathcal{A}^2} \sum_{\k_1,\k_2} (\alpha^{(i)}_0 \alpha^{(i)}_{\k} \lvert \alpha^{(j)}_{\k_2} \rvert^2 + \alpha^{(i)}_0 \alpha^{(i)}_{\k_1} \alpha^{(j)}_{\k_1} \alpha^{(j)}_{\k_2} + \lvert \alpha^{(i)}_{\k_1} \rvert^2 \alpha^{(j)}_{\k_2} \alpha^{(j)}_0 + \alpha^{(i)}_{\k_1} \alpha^{(i)}_{\k_2} \alpha^{(j)}_{\k_2} \alpha^{(j)}_0)\\ 
  &+ g\frac{N-1}{N} \frac{1}{\mathcal{A}^3} \sum_{\k_1,\k_2,\k_3} ( \lvert \alpha^{(i)}_{\k_1} \rvert^2 \alpha^{(j)}_{\k_2} \alpha^{(j)}_{\k_3} + 2\alpha^{(i)}_{\k_1} \alpha^{(i)}_{\k_2} \alpha^{(j)}_{\k_2} \alpha^{(j)}_{\k_3} + \alpha^{(i)}_{\k_1} \alpha^{(i)}_{\k_2} \lvert \alpha^{(j)}_{\k_3} \rvert^2)\,.
\end{align}

We use the equations of motion~\eqref{eq:var1X} to simplify these expressions and eliminate corrections to $U_{ij}$ of order $1/\mathcal{A}$ to obtain the following:
\begin{align}
U_{ij}(1+\delta_{ij}) &= (\omega_i+\omega_j)\frac{1}{n}\left[(1-\lvert\alpha_0^{(i)}\rvert^2)(1-\lvert\alpha_0^{(j)}\rvert^2)+(\delta_{ij}-\alpha^{(i)}_0 \alpha^{(j)}_0)^2\right]\nn\\
&-\frac{1}{n}\frac{1}{\mathcal{A}^2}\sum_{\k_1\k_2}( \frac{k_1^2+k_2^2}{2m_{\rm r}} + 2\pi n(m_{\rm X}^{-1} - m_{\rm f}^{-1}))(\lvert \alpha^{(i)}_{\k_1} \rvert^2 \lvert \alpha^{(j)}_{\k_2} \rvert^2 + \alpha^{(i)}_{\k_1} \alpha^{(i)}_{\k_2} \alpha^{(j)}_{\k_1} \alpha^{(j)}_{\k_2})\nn\\
&-\frac{2g}{\sqrt{n}}\frac{1}{\mathcal{A}^2}\sum_{\k_1\k_2}(\alpha^{(i)}_0 \alpha^{(i)}_{\k} \lvert \alpha^{(j)}_{\k_2} \rvert^2 + \alpha^{(i)}_0 \alpha^{(i)}_{\k_1} \alpha^{(j)}_{\k_1} \alpha^{(j)}_{\k_2} + \lvert \alpha^{(i)}_{\k_1} \rvert^2 \alpha^{(j)}_{\k_2} \alpha^{(j)}_0 + \alpha^{(i)}_{\k_1} \alpha^{(i)}_{\k_2} \alpha^{(j)}_{\k_2} \alpha^{(j)}_0)\nn\\
&-\frac{g}{n}\frac{1}{\mathcal{A}^3}\sum_{\k_1,\k_2,\k_3} ( \lvert \alpha^{(i)}_{\k_1} \rvert^2 \alpha^{(j)}_{\k_2} \alpha^{(j)}_{\k_3} + 2\alpha^{(i)}_{\k_1} \alpha^{(i)}_{\k_2} \alpha^{(j)}_{\k_2} \alpha^{(j)}_{\k_3} + \alpha^{(i)}_{\k_1} \alpha^{(i)}_{\k_2} \lvert \alpha^{(j)}_{\k_3} \rvert^2)\,.
\end{align}
Evaluating this expression numerically using the results of the equations of motion~\eqref{eq:var1X} yields the dashed lines shown in Fig.~\ref{fig:seaEIS}.
\end{widetext}

\bibliography{MDCS}
\end{document}